\documentclass[3p,12pt]{elsarticle}
\usepackage{indentfirst}
\usepackage{color,graphicx}
\usepackage{subfigure}
\usepackage{threeparttable}
\usepackage{array}

\usepackage{multirow}
\usepackage{amsmath}
\usepackage{soul}
\allowdisplaybreaks[4]	
\usepackage{amssymb}
\usepackage{bm}
\usepackage{hyperref}
\usepackage{lineno}

\biboptions{numbers,sort&compress}	

\journal{X}

\begin{document}

\begin{frontmatter}

\title{Wetted-Area Minimum and Inlet-Outlet Reciprocity in Optimal Manifolds of Rarefied Gas Flows}

\author{Ruifeng Yuan and Lei Wu~\corref{cor1}}
\cortext[cor1]{ Corresponding author: wul@sustech.edu.cn (Lei Wu). }

\address{Department of Mechanics and Aerospace Engineering, Southern University of Science and Technology, Shenzhen 518055, China}

\begin{abstract}
While flow optimization has been extensively studied in the continuum regime, its extension to rarefied gas flows remains less explored. Here, based on the Boltzmann model equation, an adjoint topology optimization method is employed to design two-dimensional single-inlet/multi-outlet manifolds, aiming to maximize the total mass flow rate while maintaining outflow uniformity. Two key findings are revealed. First, analogous to the Knudsen minimum in mass flow rate in the transition regime, a wetted-area minimum is identified, but in the slip flow regime. This phenomenon arises from the competition between flow bend loss and surface friction loss, with the latter being affected by velocity slip at the solid surface. Second, the inlet–outlet reciprocity emerges in the free-molecular flow regime, where the optimal design becomes invariant to inlet–outlet orientation and pressure ratio. Additional insights are gained regarding the channel curvature, compressibility effects, and the constraint of outflow uniformity.
These findings elucidate the mechanisms governing rarefied gas transport and offer design guidance for manifolds operating in vacuum environments.
\end{abstract}
\begin{keyword}
	Rarefied gas flow \sep 
	Topology optimization \sep 
	Adjoint method \sep 
	Manifold 
\end{keyword}

\end{frontmatter}

\section{Introduction}\label{sec:intro}

A manifold is a structure that distributes fluid from a single passage into multiple channels, or vice versa. Such structures are common in piping systems and widely used in internal combustion engines \cite{ceviz2010design}, heat exchangers \cite{bassiouny1984flow2}, chemical processing equipment \cite{johnson2012model}, microfluidic devices \cite{roman2007fully}, and so on. The geometric design of a manifold plays a critical role in determining flow distribution among branch channels and strongly affects the overall pressure drop, thereby influencing the performance of the associated system. Consequently, manifold design represents a classical and important problem in fluid mechanics.

Researchers have developed both theoretical and numerical frameworks to optimize manifolds. For instances,  based on simplified models that neglected wall friction in the inlet and outlet conduits, Bassiouny and Martin \cite{bassiouny1984flow} derived analytical expressions for flow distribution and pressure drop  in the manifolds of plate heat exchangers.
Pan \textit{et al.}~\cite{pan2008optimal} drew an analogy between flow resistance and electrical resistance—treating pressure drop and flow rate as voltage and current, respectively—and reformulated manifold optimization as a circuit design problem.
Emerson \textit{et al.}~\cite{emerson2006biomimetic} applied Murray’s law (relation between the diameters of parent and daughter branches in biological transport networks) to the optimal design of fluid manifolds, aiming to minimize energy losses while also accounting for the cost associated with maintaining the channel volume.
While theoretical analyses are limited to simplified geometries, subsequent works have therefore relied ever more heavily on computational fluid dynamics simulations to resolve complex flow features and optimize manifold designs, demonstrating notable improvements in flow uniformity and energy efficiency after optimization~\cite{choi1993effect,delsman2004microchannel,kubo2017level}. 

To date, most research on manifold optimization has been conducted under the continuum  assumption, where the flow is described by the Navier-Stokes equations with no-slip velocity boundary condition. However, in modern industries such as aerospace \cite{jensen2018topology}, semiconductor manufacturing \cite{bakshi2009euv}, and advanced instrumentation \cite{garimella2013simulation}, many piping systems involve vacuum environments, necessitating consideration of rarefied gas effects in manifold design. Specifically, for pipe/channel flow, rarefaction effects mainly cause two impacts: the velocity slip at the gas-solid interface and  the breakdown of linear stress constitutive relation. In particular, for the Poiseuille flow in micro-channel, the Navier-Stokes equations with no-slip condition predict a monotonic decrease in normalized mass flow rate (MFR) toward zero as gas rarefaction increases. In contrast, experiments and the Boltzmann equation reveal a non-monotonic trend---initially decreasing then increasing---resulting in the famous ``Knudsen minimum" phenomenon in rarefied gas dynamics \cite{ewart2007mass}. Therefore, manifold optimization of rarefied gas flows is expected to yield fundamentally different results compared to those of continuum flows. 

Nevertheless, this problem remains under-explored, primarily due to the challenges associated with the effective solution of both the Boltzmann equation and its adjoint equation~\cite{sato2019topology,caflisch2021adjoint}.
In this paper, we apply a Boltzmann-based topology-optimization framework~\cite{yuan2024design} to concurrently tailor the size, shape, and topology of 2D single-inlet/multi-outlet manifolds. 
Our results show that the optimal configurations vary non-monotonically with the Knudsen number, exhibiting a wetted-area minimum in the slip regime and inlet--outlet reciprocity in the free-molecular regime.

\section{Problem and Method}\label{sec:method}

\begin{figure}[t]
	\centering
	\includegraphics[width=0.55\textwidth]{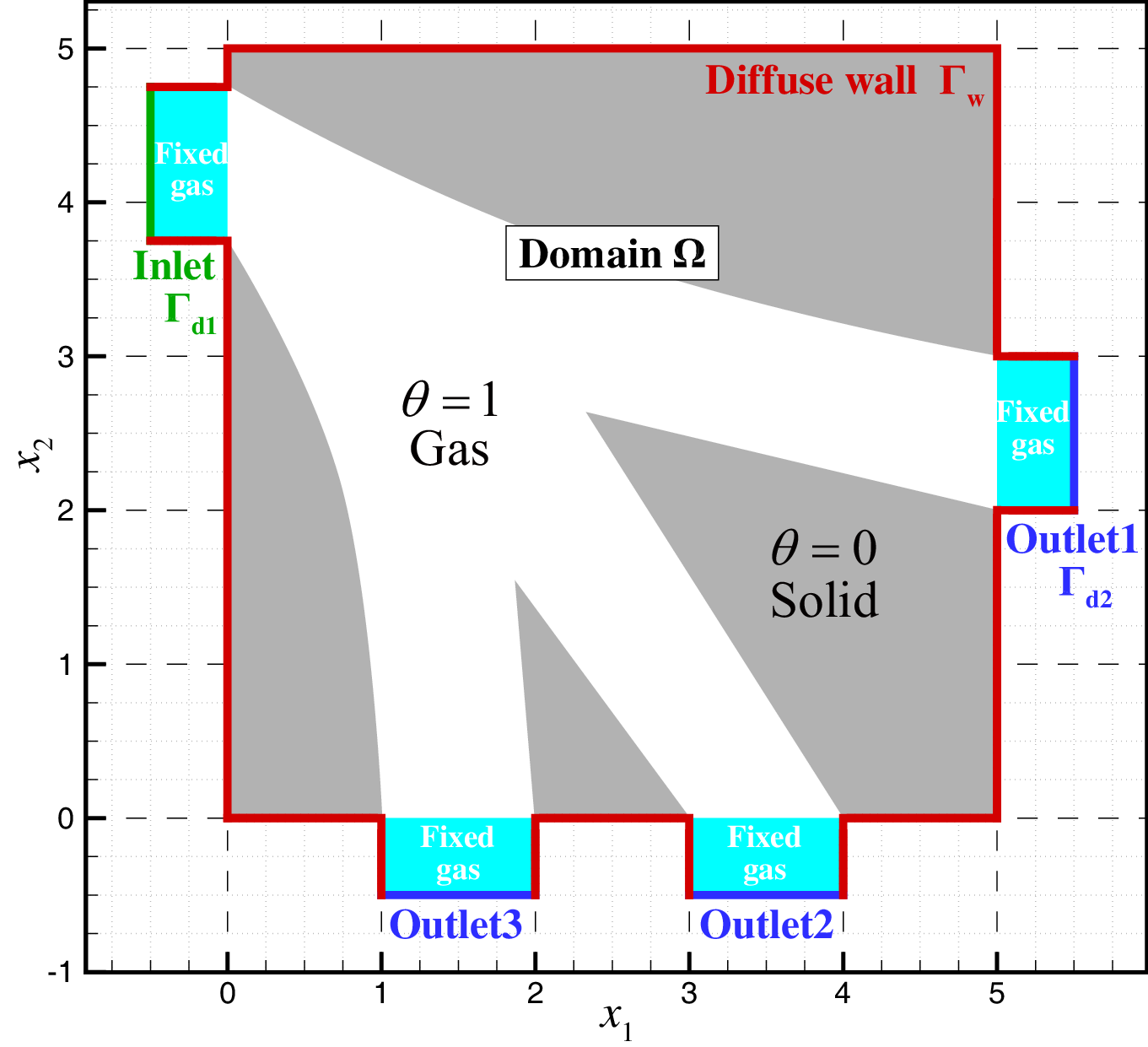}
	\caption{\label{fig:formula_domain}
		Schematic of the topology optimization of a 2D manifold based on the material density $\theta$ (the gas and solid regions are represented by $\theta=1$ and 0, respectively).
		The cyan regions are fixed as gas. $\Omega$ is the $5\times5$ square design domain. 
		The green edge $\Gamma _{\rm d1}$ is the inlet, the blue edges $\Gamma _{\rm d2}$ are three outlets, and the red edges $\Gamma _{\rm w}$ are solid walls.
		The coordinates are normalized by the inlet/outlet channel height $H$.
	}
\end{figure}

Consider the topology optimization of a 2D intake manifold shown in figure~\ref{fig:formula_domain}. 
The manifold to be designed has one inlet and three outlets with the channel height $H=1$. To prevent numerical errors at inlet/outlets from affecting optimization results, buffer regions fixed as gas are implemented near the four corresponding boundaries. The design domain $\Omega$ is a $5\times5$ square centered at $[x_1,x_2]=[2.5,2.5]$. 
The gas flow is driven by the pressure difference between the inlet and outlets, and for simplicity the Dirichlet boundary condition is enforced on the inlet and outlet boundaries ($\Gamma _{\rm d1}$ and $\Gamma _{\rm d2}$), where the corresponding velocity distribution functions are set as the Maxwellian distribution under the far-field gas states (i.e. the reservoir states) $[\rho_{\rm in},\bm 0, T_\infty]$ (density, velocity, and temperature for inlet) and $[\rho_{\rm out},\bm 0, T_\infty]$ (for outlet), with $\rho_{\rm in}/\rho_{\rm out}=100$.
On other solid boundaries, the diffuse  boundary condition is enforced with a wall temperature $T_{\rm w}=T_{\infty}$.

Following the practice of classical density-based topology optimization methods, the material density $\theta$ is used to distinguish gas ($\theta=1$) and solid ($\theta=0$) regions~\cite{bendsoe2003topology,borrvall2003topology}. The manifold optimization problem is then transformed into an optimization of the distribution of $\theta$.
The total MFRs should be maximized after optimization. Furthermore, flow uniformity among multiple outlets constitutes a critical design consideration in manifold systems. Thus, the objective function $J$ is formulated as:
\begin{equation}\label{eq:obj}
	J = -\bar F + \epsilon{S^2},
\end{equation}
where $\bar F$ and $S^2$ are respectively the average and variance of the MFRs:
\begin{equation}
	\bar F = \frac{1}{3}\sum\limits_{i = 1}^3 {{F_{{\rm{o}}i}}} ,\quad {S^2} = \frac{1}{3}\sum\limits_{i = 1}^3 {{{\left( {{F_{{\rm{o}}i}} - \bar F} \right)}^2}},
\end{equation}
with $F_{\rm o1}\sim F_{\rm o3}$ denoting the MFRs for outlet1$\sim$outlet3, and $\epsilon$ being the penalty factor. Roughly speaking, we choose $\epsilon=10$ for Kn = 0.001, and $\epsilon=300$ for Kn larger than 0.1; reasons will be given in Section~\ref{sec:intake_np}.

Moreover, the manifold has a constraint for the maximum gas volume, i.e., $V_{\max }=8$ (exclude the fixed gas regions which have a total volume of 2). Finally, the design problem is formulated as:
\begin{equation}\label{eqn:formula_opt}
\mathop {\min }\limits_{\theta \in [0,1]} ~J , \quad
{\rm{s}}{\rm{.t}}{\rm{.}}~ \int_\Omega  {\theta d\Omega }  - {V_{\max }} \le 0.
\end{equation}

\subsection{Gas-kinetic model equation}

One important parameter to indicate the level of gas rarefaction is the Knudsen number (Kn), which is the ratio of the molecular mean free path $l_{\rm mfp}$ to the characteristic length of the gas flow (here it is $H$):
\begin{equation}\label{eqn:kndefine}
	{\rm{Kn}} = \frac{l_{\rm mfp}}{H}.
\end{equation}
When Kn is very small (${\rm Kn}<0.001$ for the continuum regime), the Navier-Stokes equation with no-slip velocity condition can be used. As Kn increases, distinct flow regimes emerge: in the slip regime ($0.001<{\rm Kn}<0.1$), the velocity slip and temperature jump occur at the gas-solid interface; in the transition regime ($0.1<{\rm Kn}<10$), the linear constitutive relations break down and the Navier-Stokes equations are no longer valid; in the free-molecular regime (${\rm Kn}>10$), intermolecular collisions become negligible and the flow is governed by the collision between gas molecules and solid surface. 
The gas-kinetic equation and the diffuse boundary condition provide a uniform description of continuum and rarefied gas flows.

Without loss of generality, in this paper, the Boltzmann Bhatnagar-Gross-Krook (BGK) model equation is used to describe the rarefied gas dynamics~\cite{bhatnagar1954model}:
\begin{equation}\label{eqn:bgk_reduce}
\frac{{\partial \bm f}}{{\partial t}} + \bm v \cdot \nabla \bm f = \frac{{{\bm g } - \bm f}}{{{\tau }}}.
\end{equation}
As we focus on the 2D problem, $\bm f=[f_1,f_2]$ is the marginal velocity distribution function, $\bm v=[v_1,v_2]$ is the molecular velocity, and
$\bm g={\bm g_{\rm{M}}}(\rho ,{\bm u},{T})
={[{g_{1}},{g_{2}}]}$ is the reduced Maxwellian equilibrium distribution determined by the density $\rho$, flow velocity ${\bm u }=(u_1,u_2)$, and temperature $T$:
\begin{equation}
{g_{1}} =\frac{ \rho}{  2\pi RT}
\exp\left[ - \frac{{{{(\bm v - {\bm u })}^2}}}{{2R{T }}}\right],
\quad
{g_{2}} = \frac{{{R{T } }}}{2}g_{1},
\end{equation}
where $R$ is the specific gas constant.
The relaxation time is $\tau=\mu / p$, with the dynamical viscosity coefficient $\mu \propto \sqrt T $ and the gas pressure $p=\rho RT$. Then the mean free path $l_{\rm mfp}$ is determined by the
hard-sphere model \cite{Bird1994Molecular}:
\begin{equation}\label{eqn:mfpdefine}
l_{\rm mfp}= \frac{{16}}{5}\tau\sqrt {\frac{{RT}}{{2\pi }}} .
\end{equation}

The macroscopic gas state can be obtained from the moments of  velocity distribution function, e.g., $\rho=\int f_1 d\Xi$, $\rho\bm{u}=\int \bm{v} f_1 d\Xi$, and $\frac{1}{2}\rho \bm u ^2 + \frac{3}{2}\rho R T =\frac{1}{2}\int \bm{v}\cdot\bm{v} f_1 d\Xi+\int f_2 d\Xi$, where $d\Xi  = dv_1dv_2$ is the molecular velocity space element.

The kinetic equation~\eqref{eqn:bgk_reduce} governs the flow in the bulk region. On the boundary of the computational domain, we consider two types of boundary conditions. The first is the Dirichlet boundary condition for the far-field free-stream boundary $\Gamma _{\rm d}$:
\begin{equation}\label{eqn:bc_drlt}
	\bm f = \bm f_{\rm d} \quad {\rm{in}}\quad {\Gamma _{\rm d}} \times {\Xi ^ - },
\end{equation}
where ${\Xi ^ \pm } = \left\{ \bm v | \bm v \cdot \bm n \gtrless 0 \right\}$ representing the molecular velocity moving out of or into the boundary with a normal unit vector $\bm n$ pointing outward from the computational domain, and $\bm f_{\rm d}$ is a given fixed velocity distribution which is normally set as the Maxwellian distribution of the far-field condition. The second is the diffuse boundary condition for the stationary solid wall $\Gamma _{\rm w}$: 
\begin{equation}\label{eqn:formula_bgk_fdw0}
	\bm f ={\bm g_{\rm{M}}}({\rho _{\rm{w}}},\bm 0,T_{\rm{w}}), \quad {\rm{in}}\quad {\Gamma _{\rm w}} \times {\Xi ^ - },
\end{equation}
where $T_{\rm{w}}$ is a prescribed wall temperature; $\rho _{\rm{w}}$ is determined by the zero mass flux condition on the solid wall and can be written as
\begin{equation}\label{eqn:formula_bgk_fdw}
	{\rho _{\rm{w}}} = \sqrt {\frac{{2\pi }}{{R{T_{\rm{w}}}}}} \int_{{\Xi ^ + }} {\bm v \cdot \bm n{f_1}d\Xi } .
\end{equation}

\subsection{The fictitious porosity model in topological optimization}\label{sec:primal}

To apply the Boltzmann BGK equation to the topology optimization problem, a fictitious porosity model is constructed to exert force on the flow field based on the value of $\theta$, thereby simulating the obstruction of gas flow by solid material. That is, 
although realistically the material density $\theta$ should be either 1 (gas) or 0 (solid), due to the difficulty in handling such a discontinuous distribution of $\theta$ in numerical treatment, we allow the material to continuously vary between $\theta=1$ (pure gas) and $\theta=0$ (pure solid).

The equilibrium state $\bm g$ and the relaxation time $\tau$ of the original BGK equation~\eqref{eqn:bgk_reduce} is modified to fictitious quantities $\bm g_\theta$ and $\tau _\theta$. 
Here $\bm g_\theta={\bm g_{\rm{M}}}(\rho ,{\bm u_\theta },{T_\theta })$ is the Maxwellian distribution determined by the fictitious macroscopic state $[\rho, {\bm u_\theta }, {T_\theta }]$, where $\rho$ is the original gas density but the fictitious velocity and temperature are calculated by~\cite{yuan2024design}
\begin{equation}
{\bm u_\theta } = {\theta _w}\bm u, \quad
{T_\theta } = {\theta _w}T + \left( {1 - {\theta _w}} \right){T_s},
\end{equation}
where $T_s$ is the prescribed solid temperature, and $\theta _w$ is related to the material density $\theta$ as
\begin{equation}\label{eqn:q_w}
{\theta _w}  = \left\{ \begin{aligned}
& \frac{{{q_w}\theta }}{{1 + {q_w} - \theta }},\quad q_w >  0\\
& \frac{{\left( {1 - {q_w }} \right)\theta }}{{\theta - q_w}},\quad q_w <  0
\end{aligned} \right. 
\end{equation}
So for $\theta=1$ (pure gas), we have $\theta_w=1$, and the macroscopic state for the equilibrium distribution $\bm g_\theta$ is just the original gas state $[\rho, {\bm u }, {T }]$. When $\theta=0$ (pure solid), it is $[\rho, {\bm 0 }, {T_s }]$.
The parameter $q_w$ determines how fast the variables $\bm u_\theta, T_\theta$ vary with $\theta$, which will influence the quality and efficiency of topology optimization (details in Ref.~\cite{yuan2024design}). Here we choose $q_w=-0.1$ for Kn = 0.001, $q_w=0.01$ for Kn = 0.1, and $q_w=0.001$ for Kn = 10. 

For the fictitious relaxation time $\tau _\theta$, it is calculated as
\begin{equation}
\frac{1}{{{\tau _\theta }}} = \frac{\theta _\tau }{\tau } + \frac{1 - {\theta _\tau }} {{{\tau _s}}},
\end{equation}
where $\theta _\tau$ is related to the material density $\theta$ by
\begin{equation}\label{eqn:q_tau}
{\theta _\tau } = \frac{{\left( {1 + 0.005} \right)\theta }}{{0.005 + \theta }}.
\end{equation}
So for $\theta=1$ we have $\tau _\theta=\tau$ and recover exactly the original BGK equation \eqref{eqn:bgk_reduce}, while for $\theta=0$ we have $\tau _\theta=\tau _s$. $\tau _s$ is set to a very small value to ensure that when $\theta=0$ the gas can quickly relax to the stationary solid state with a Maxwellian distribution, so that the Maxwell diffuse condition is realized at the gas-solid interface. Here $\tau _s$ is equal to $0.1$ times the local CFL time step (see Ref.~\cite{yuan2024design} for details).

\subsection{Sensitivity analysis}\label{sec:sens}

Since simulating rarefied gas flows is quite expensive, we have to perform the optimization by the gradient-based optimizer, which requires calculating the derivative (or sensitivity) of the objective $J$ with respect to $\theta$. Here the sensitivity is obtained by the continuous adjoint method \cite{peter2010numerical}. A detailed analysis process is provided in Ref.~\cite{yuan2024design}, and we will not extensively discuss it in this paper, but only present the final equations. 

First we should solve the adjoint variable $\bm \phi=[\phi_1,\phi_2]$ from the following adjoint governing equation:
\begin{equation}\label{eqn:formula_adjointbgk}
\left.
\begin{aligned}
 - \bm v \cdot \nabla \bm \phi  = \frac{{{\bm \phi _{{\rm{eq}}}} - \bm \phi }}{{{\tau _\theta }}} + {\bm \phi _\tau } + {\bm \phi _J}, \quad {\rm{in}}\quad \Omega_C  \times \Xi, \\
\bm \phi  = \bm 0, \quad {\rm{in}} \quad {\Gamma _{\rm d}} \times {\Xi ^ + },\\
\bm \phi  = \left( \begin{aligned}
 - \sqrt {\frac{{2\pi }}{{R{T_{\rm{w}}}}}} \int_{{\Xi ^ - }} \bm v & \cdot \bm n \bm\phi  \cdot {\bm g_{\rm{M}}}(1,\bm 0,T_{\rm{w}})d\Xi  \\
 0
\end{aligned} \right),\quad {\rm{in}}\quad {\Gamma _{\rm w}} \times {\Xi ^ + }.
\end{aligned}
\right\}
\end{equation}
Note that we only consider steady problem so the temporal term is omitted. 
$\Omega_{\rm C}$ is the whole computational domain. The right-hand-side terms are
\begin{equation}\label{eqn:adjoint_eqdefines}
{\bm \phi _{\rm eq}} = \hat {\bm W} \cdot {\frac{{\partial {\bm W_\theta }}}{{\partial \bm W}}} \cdot {\bf{\Psi }},\quad {\bm \phi _\tau } =  - {{\hat \rho }_\tau }\frac{1}{{{\tau _\theta }}}\frac{{\partial {\tau _\theta }}}{{\partial \bm W}} \cdot {\bf{\Psi }},
\end{equation}
with
\begin{equation}\label{eqn:adjoint_macdefines}
\hat {\bm W} = \int_\Xi  {\bm \phi  \cdot \frac{{\partial {\bm g_\theta }}}{{\partial {\bm W_\theta }}}d\Xi } ,  \quad {\hat \rho _\tau } = \int_\Xi  {\bm \phi  \cdot \frac{{{\bm g_\theta } - \bm f}}{{{\tau _\theta }}}d\Xi }.
\end{equation}
$\bm W=[\rho,\rho\bm u,\frac{1}{2}\rho \bm u ^2 + \frac{3}{2}\rho R T ]$ are the macroscopic conservative variables and $\bm W_\theta$ correspond to the variables calculated by $\bm u_\theta$ and $T_\theta$. $\bf{\Psi }$ is the moments tensor
\begin{equation}
{\bf{\Psi }} = \left[ {\begin{array}{*{20}{c}}
\begin{array}{l}
1\\
\bm v\\
{\textstyle{1 \over 2}}{\bm v^2}
\end{array}&\begin{array}{l}
0\\
0\\
1
\end{array}
\end{array}} \right].
\end{equation}

The term ${\bm \phi _J}$ is related to the objective $J$. Specifically, in this study $J$ is determined from the MFRs $F_{\rm o1}\sim F_{\rm o3}$ at outlet1$\sim$outlet3, which are calculated by
\begin{equation}
{F_{{\rm o}i}} = \frac{1}{{\Delta x}}\int_{{\Omega _{{\rm o}i}}}^{} {\rho \bm u \cdot \bm nd\Omega } ,
\end{equation}
where $\Omega _{{\rm o}i}$ represents the region occupied by the first layer of mesh cells adjacent to the $i$-th outlet, and $\Delta x$ is the mesh cell height.
Then we have
\begin{equation}
{{\bm \phi }_J} = \left\{ 
\begin{aligned}
 - \frac{1}{{\Delta x}}\frac{{\partial J}}{{\partial {F_{{\rm{o}}i}}}}\left[ {\bm v \cdot \bm n,0} \right],\quad {\rm{in}}\quad {\Omega _{{\rm o}i}},\\
0,\quad {\rm{in}}\quad {\Omega _{\rm C}}\backslash {\Omega _{{\rm{o}}i}}.
\end{aligned} 
\right.
\end{equation}

After obtaining the adjoint variable $\bm \phi$, the sensitivity with respect to the material density $\theta$ can be calculated as
\begin{equation}\label{eqn:sens_theta}
 d{J}= \int_\Omega  \left(  - \frac{1}{{{\tau _\theta }}}\hat {\bm W}\frac{{\partial {\bm W_\theta }}}{{\partial \theta }}{\rm{ + }}{\hat \rho _\tau }\frac{1}{{{\tau _\theta }}}\frac{{\partial {\tau _\theta }}}{{\partial \theta }}\right)
 \delta \theta d\Omega .
\end{equation}

\subsection{Optimization procedure}\label{sec:proc}

The solution of the design problem \eqref{eqn:formula_opt} follows the computational procedure of the general gradient-based optimization method, which is outlined below:
\begin{enumerate}
\item Solve the gas-kinetic governing equation to obtain the objective $J$. Here we adopt a multiscale numerical scheme \cite{yuan2020conservative, yuan2021multi} to ensure good efficiency and accuracy.
\item Calculate the derivative (or sensitivity) of the objective $J$ with respect to the material density $\theta$. This is accomplished by using the adjoint method introduced in Section \ref{sec:sens}.
\item Substitute the objective and sensitivity into the optimizer to optimize the distribution of $\theta$. Here we employ the algorithm of moving asymptotes  \cite{svanberg1987method,svanberg2002class}  implemented in the NLopt library of Johnson \cite{johnson2007nLopt}.
\end{enumerate}
Note that, for the sake of readability, the above computational procedure omits many detailed steps (e.g., filtering and projection operations applied to $\theta$). A more comprehensive description of the computational procedure can be found in our recent work~\cite{yuan2024design}.

\section{Numerical results}\label{sec:test}

In this section, we first optimize the 2D manifold to investigate the effect of gas rarefaction on the optimal configuration. Then, based on the insights gained in understanding the optimization, we explore the effect of exchanging inlet and outlet conditions. Finally, we study the impact of flow uniformity constraint on the optimal configuration.

\subsection{Simulation parameters}

For the mesh discretization, in the case ${\rm Kn}=0.001$, the computational domain is discretized by $200 \times 200$ uniform cells. In the molecular velocity space, a $20 \times 20$ uniform discretization is adopted for a square domain within the velocity range $[ - 6{a_\infty },6{a_\infty }]$ in both $v_1,v_2$ directions, where $a_\infty$ is the acoustic velocity of the inlet gas state. 
In the cases of ${\rm Kn}=0.1, 10$, due to the increase of mean free path, a coarser physical space mesh with the mesh size of $0.05$ ($100 \times 100$ cells for the square design domain), and a refined velocity space mesh with 1628 nonuniform cells are adopted \cite{yuan2020conservative}.
According to our mesh dependency studies, the above mesh densities are fine enough for the current optimization problem under the considered flow conditions.

\begin{figure}[p]
	\centering
	\subfigure[\label{fig:case1a}Optimized configuration, with two flow bifurcation points indicated by $B_1,B_2$. $O_1,O_2$ and $O_3$ are central points of outlet boundaries. $S$ is the point to measure slip velocity.]{
		\includegraphics[width=0.33\textwidth]{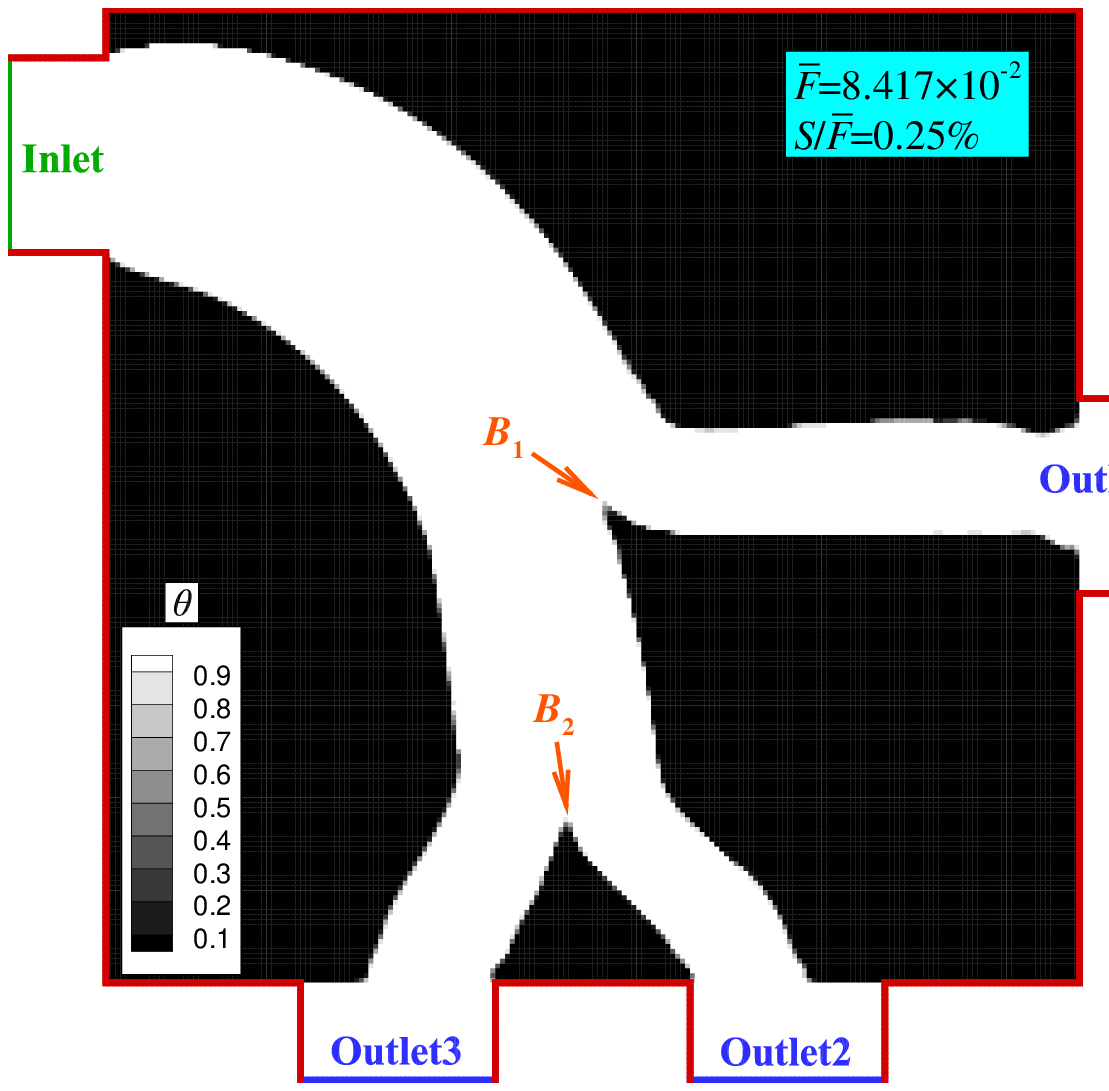}
		\includegraphics[width=0.33\textwidth]{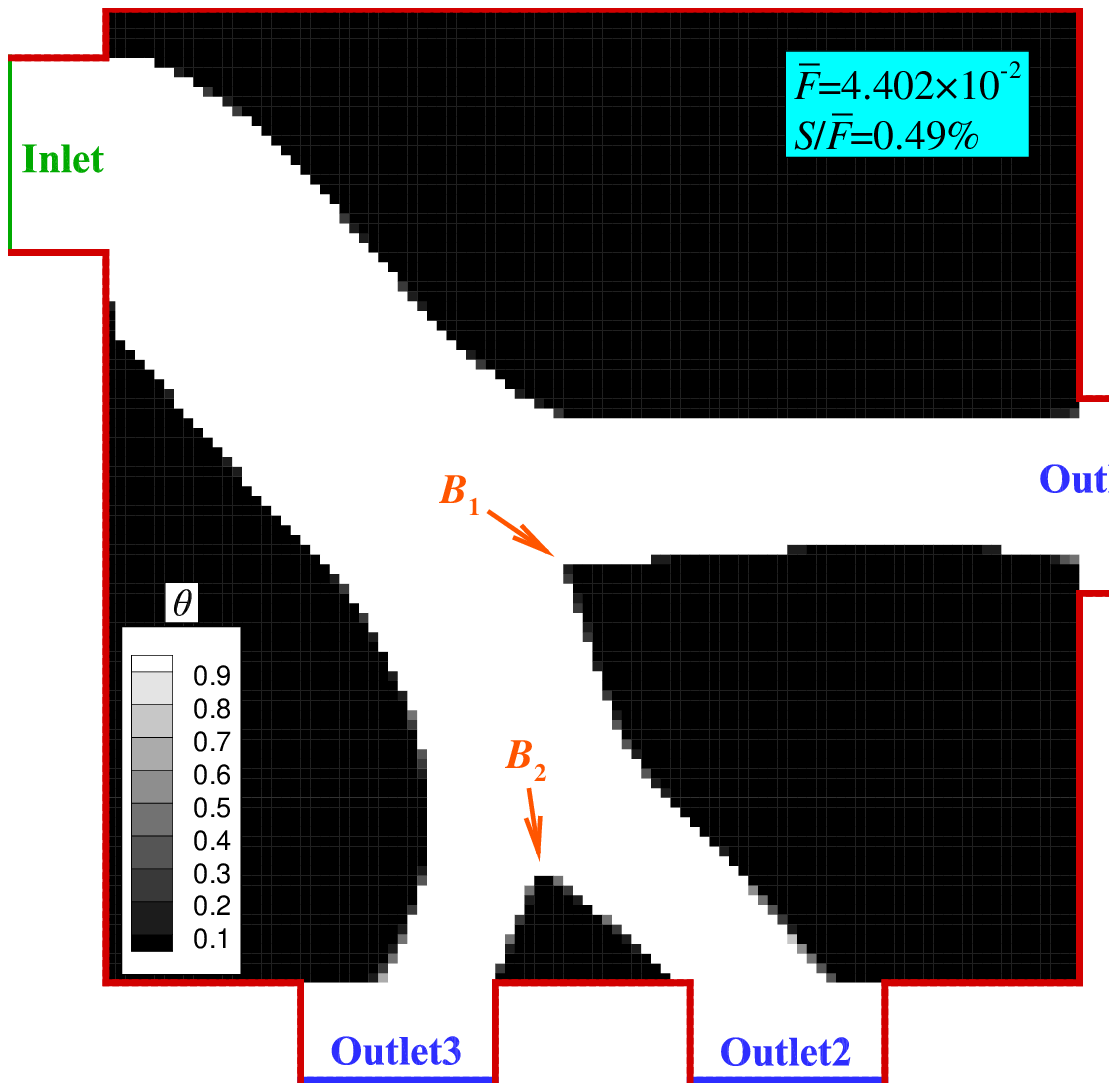}
		\includegraphics[width=0.33\textwidth]{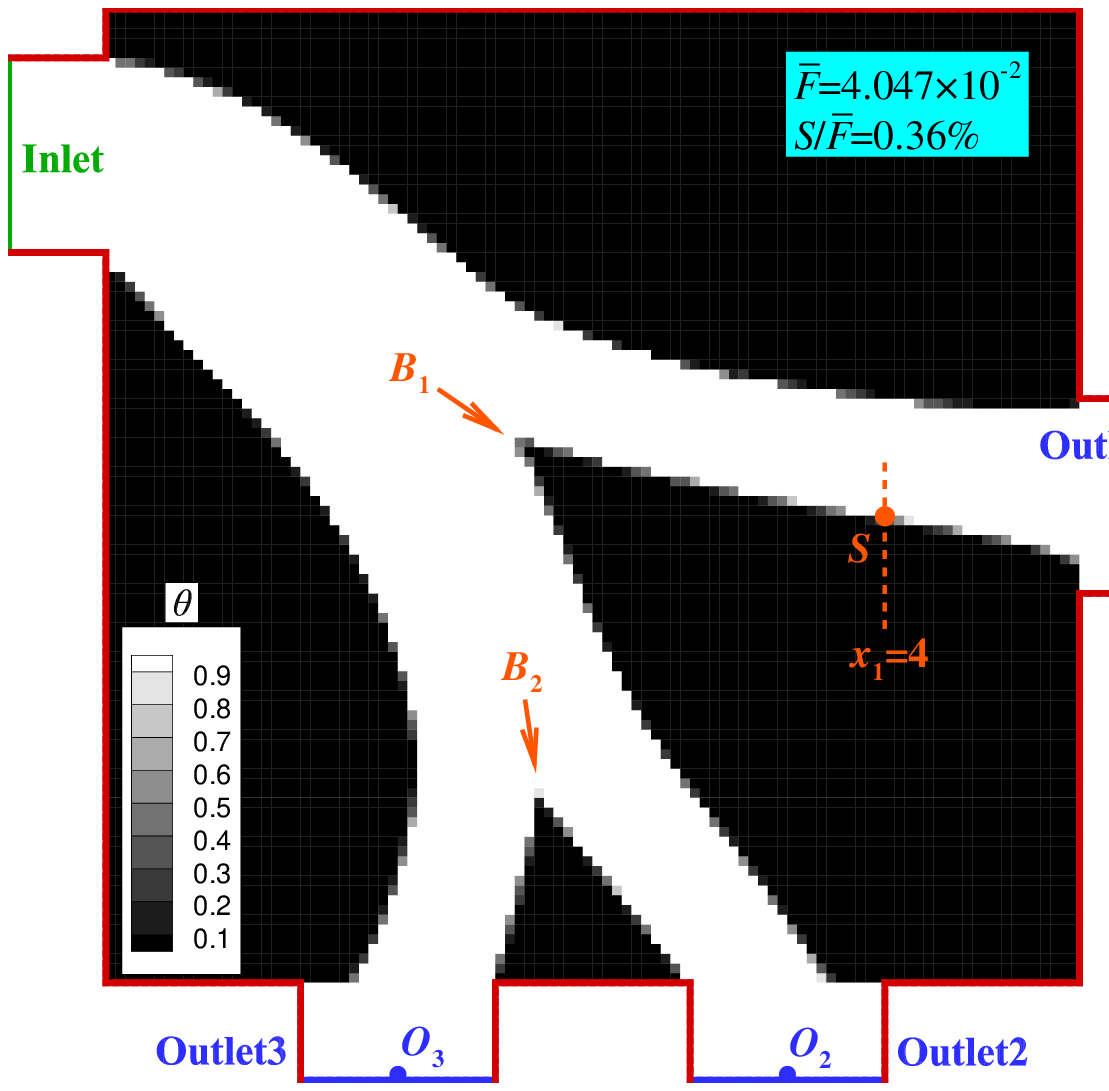}
	}
	\subfigure[\label{fig:case1b}Distribution of Mach number]{
		\includegraphics[width=0.33\textwidth]{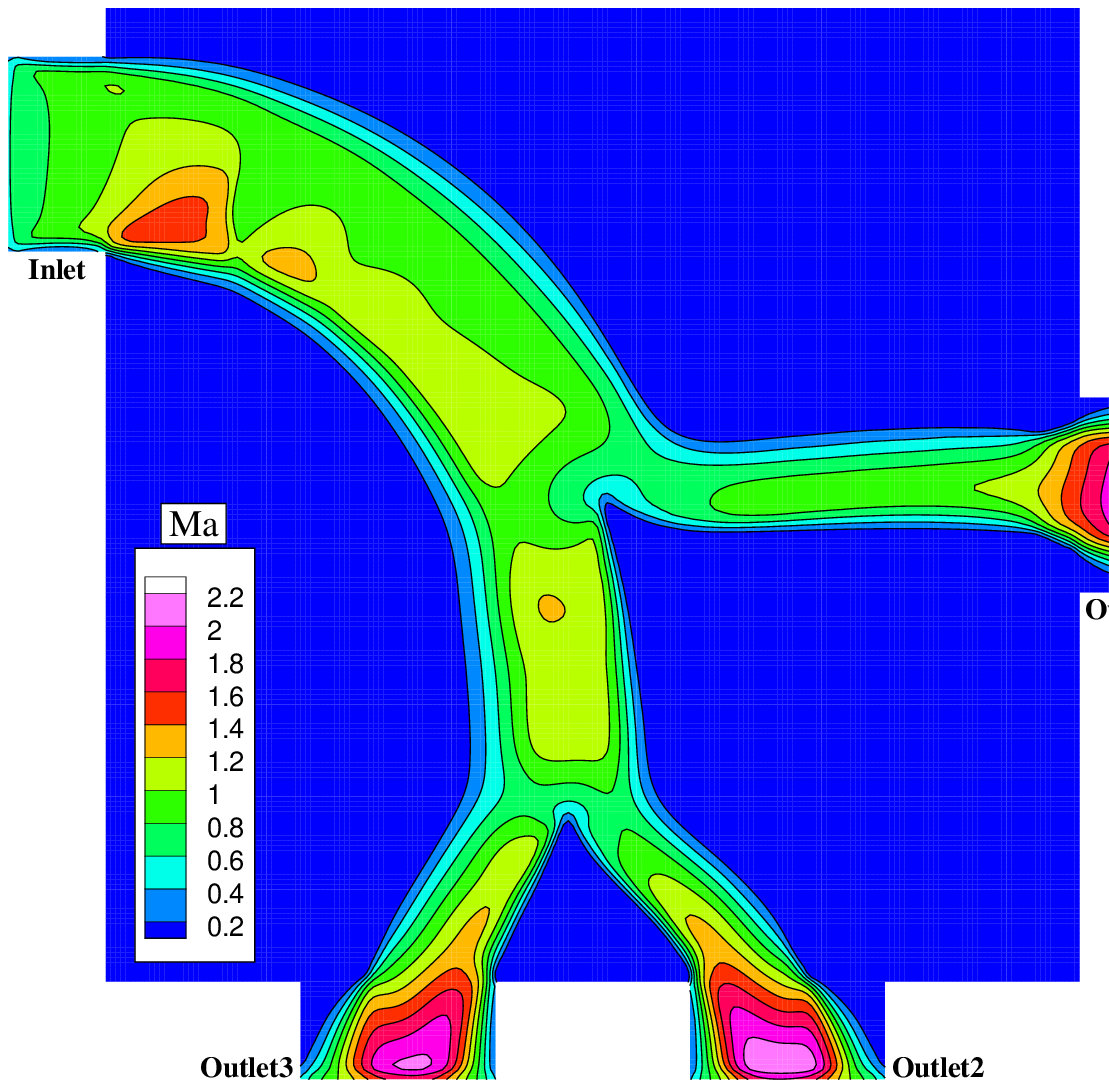}
		\includegraphics[width=0.33\textwidth]{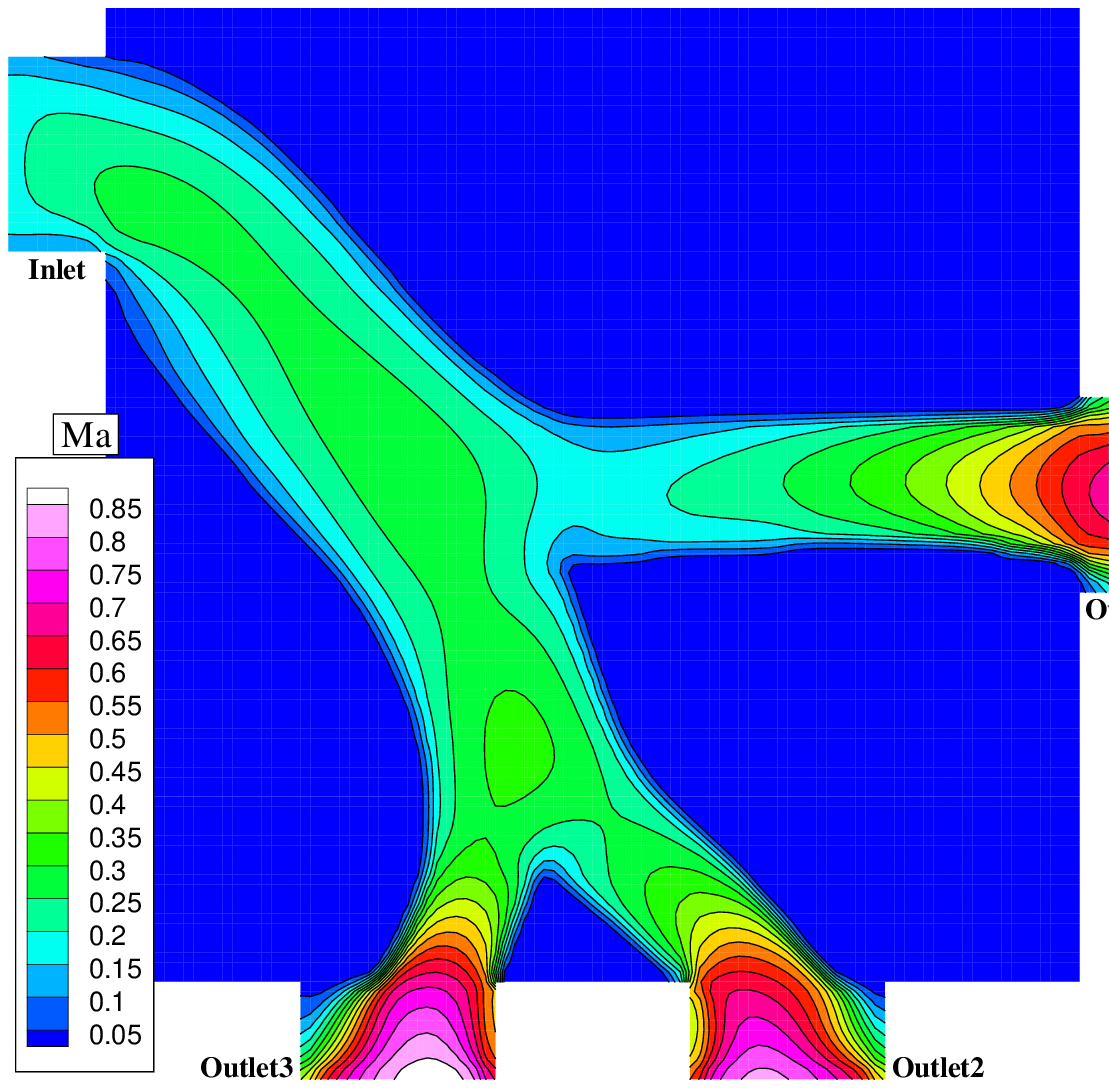}
		\includegraphics[width=0.33\textwidth]{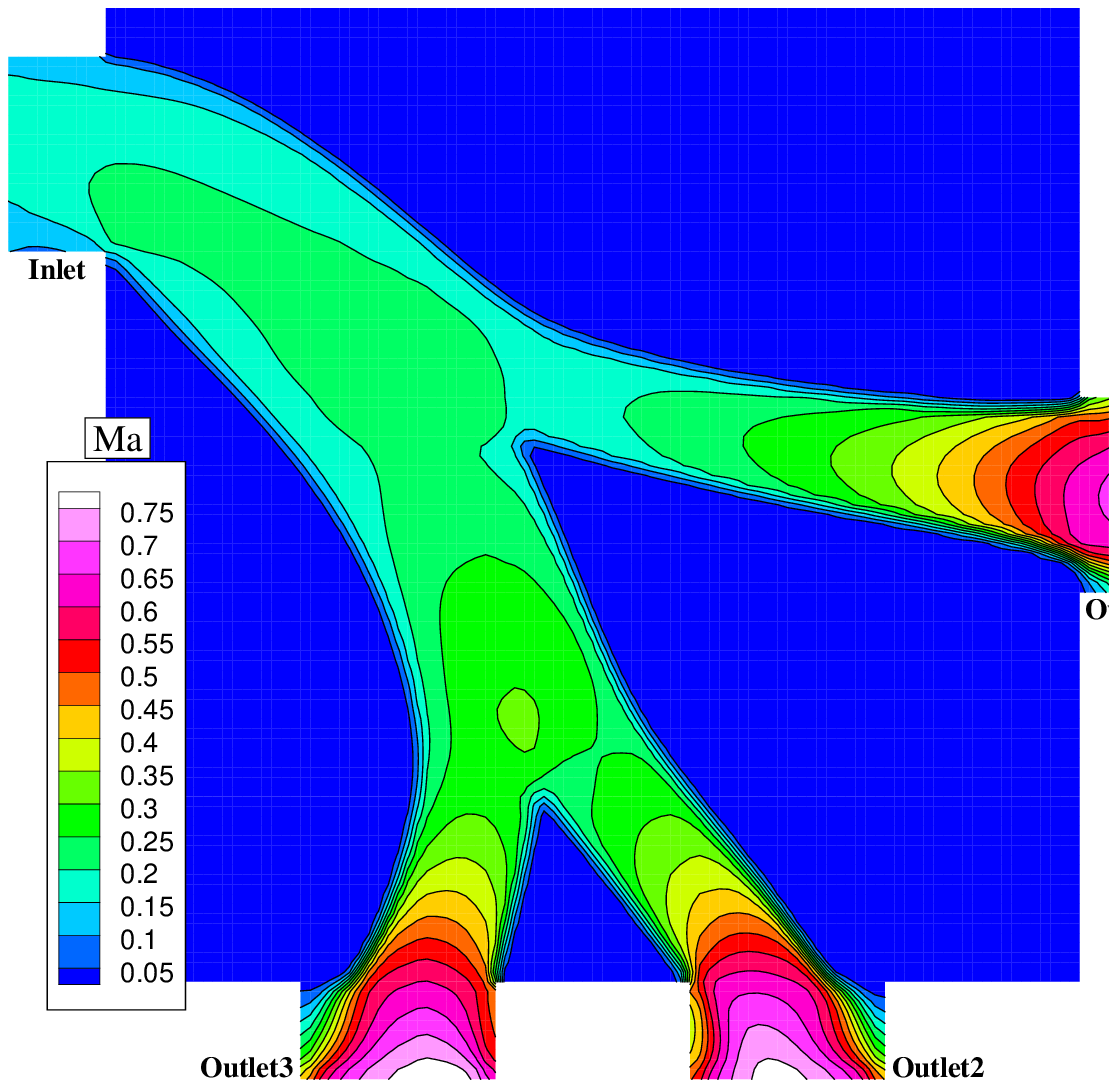}
	}
	\subfigure[Streamlines and pressure contours]{
		\includegraphics[width=0.33\textwidth]{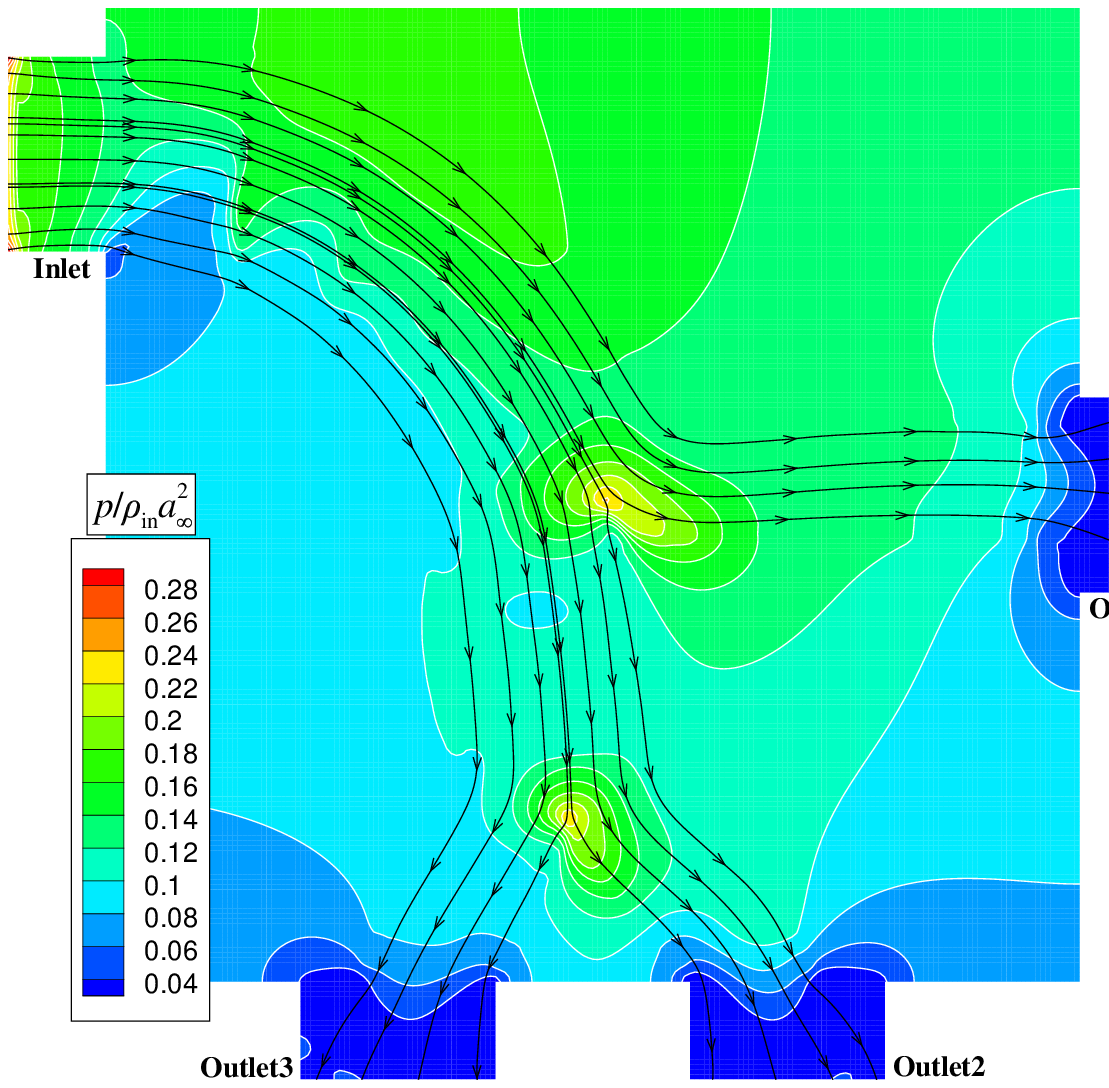}
		\includegraphics[width=0.33\textwidth]{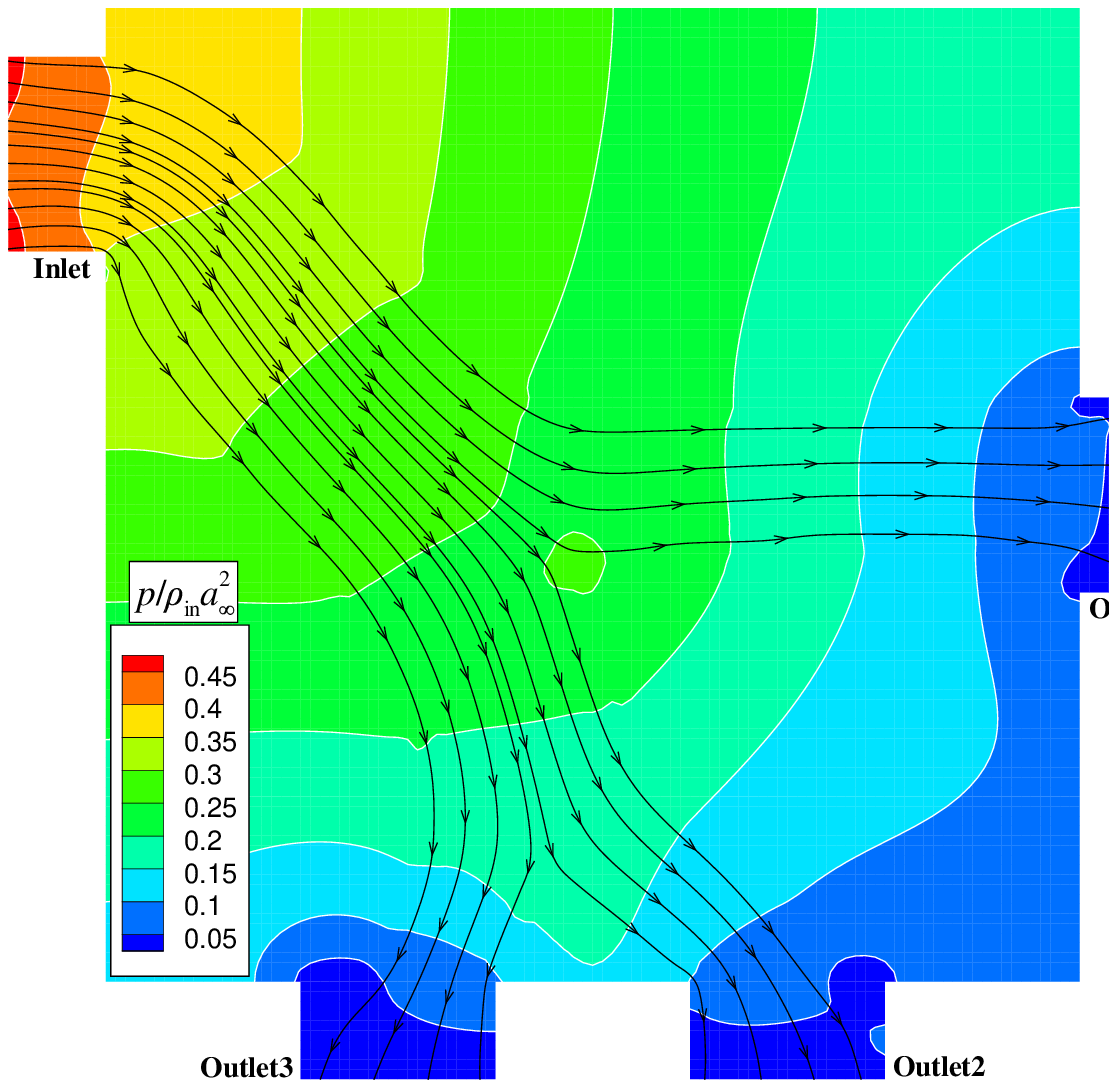}
		\includegraphics[width=0.33\textwidth]{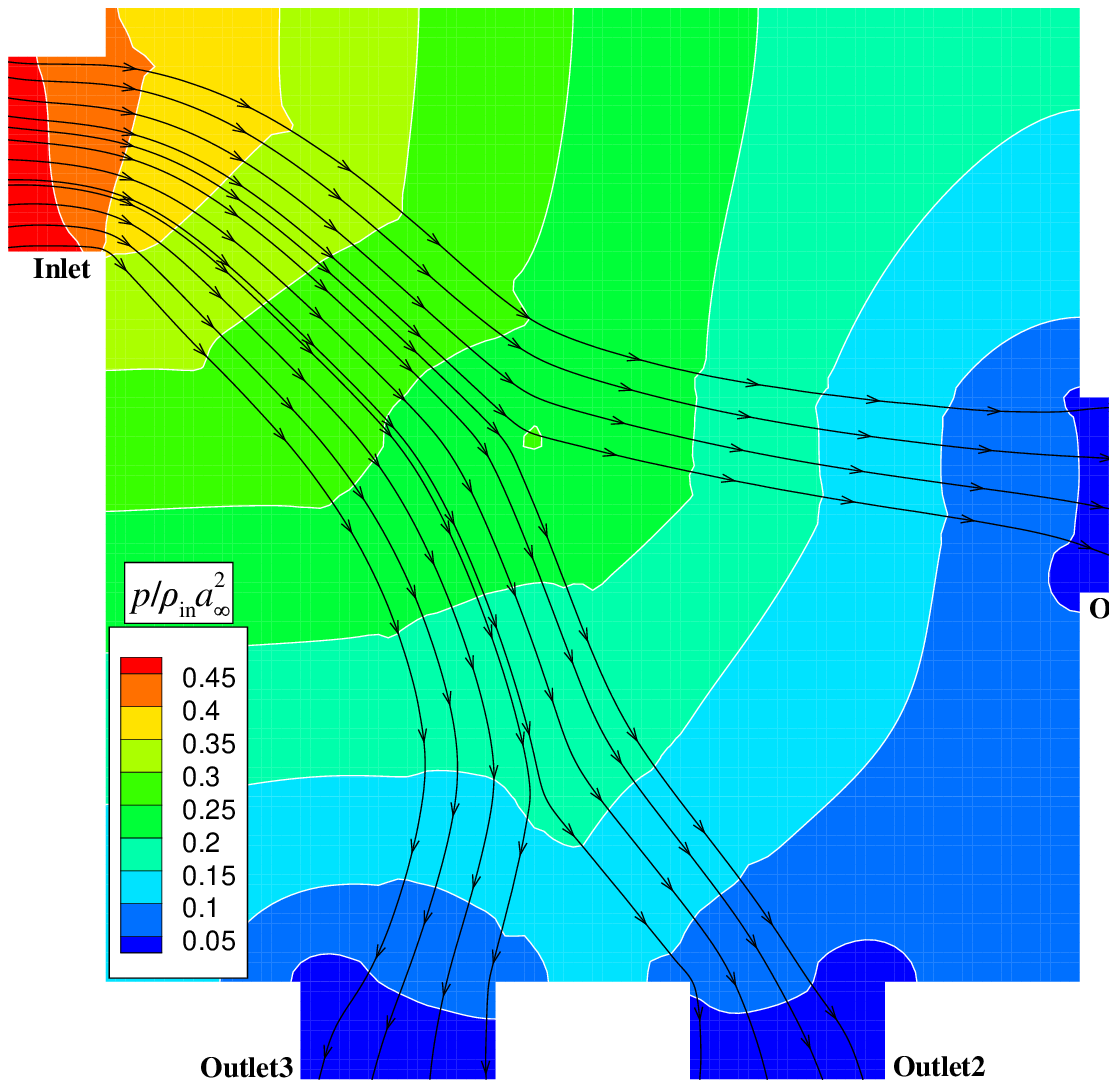}
	}
	\caption{\label{fig:case1}Optimization of a 2D intake manifold for MFR under flow uniformity constraint. From left to right are the results for ${\rm Kn}=0.001,0.1,$ and 10, respectively.
	}
\end{figure}

\subsection{Case 1: Optimization of intake manifold for MFR under flow uniformity constraint}\label{sec:intake}

The results of the manifold optimization for MFR under the flow uniformity constraint are presented in figure~\ref{fig:case1} for three representative Knudsen numbers: 0.001, 0.1, and 10. The MFRs, which are normalized by $\rho_{\rm in} a_\infty H$, are $8.417\times10^{-2}$, $4.402\times10^{-2}$, and $4.047\times10^{-2}$, respectively.
The small coefficient of variation, $S/\bar{F}$, indicated in the legend of figure~\ref{fig:case1a}, demonstrates that excellent flow uniformity is achieved as expected. Although the optimal manifolds at different Kn share broadly similar topologies, they exhibit distinct detailed configurations. These differences are analyzed in the following three aspects.

\textbf{Curvature of channel}: As shown in figure~\ref{fig:case1a}, at small Knudsen numbers, the optimization tends to produce channels with noticeable curvature, whereas under rarefied conditions, straighter channels are preferred. This trend is consistent with previous studies on single-channel optimization~\cite{gersborg2005topology,pingen2007topology,sato2019topology}. The underlying mechanism  is that, inertial effects dominate the gas flow at small Knudsen numbers, where an abrupt change in flow direction causes significant momentum loss \cite{pingen2007topology} and may even induce flow separation. Therefore, employing channels with moderate curvature to smoothly alter flow direction is advantageous at small Knudsen numbers. 
In contrast, at large Knudsen numbers where molecules tend to move ballistically, using straight channels shortens the flow path and increases the channel width (due to the volume constraint), thereby reducing losses associated with wall friction.

\textbf{Effect of compressibility}: Figure~\ref{fig:case1}(b, c) illustrates that for the rarefied cases, 
Kn = 0.1 and Kn = 10, the gas flows remain entirely subsonic, with pressure gradually decreasing along the flow direction and pressure contours oriented nearly perpendicular to the streamlines.
In contrast, at Kn=0.001, owing to the low viscosity, the gas accelerates to supersonic speeds multiple times, with several shock waves observed in the bend regions and upstream of the bifurcation points. The flow eventually becomes fully supersonic near the outlets. Correspondingly, the optimized channel near the inlet resembles a diffuser, while classical Laval nozzle structures are clearly visible near the outlets, facilitating acceleration of the gas to supersonic velocities.


Additionally, at Kn = 0.001, the flow in the bend section is strongly affected by centrifugal forces arising from the high gas velocity and channel curvature. The pressure gradient is nearly aligned with the outward normal direction of the curvature. Under the combined effects of centrifugal force and compressibility, the gas density on the inner side of the bend becomes lower than that on the outer side. Consequently, to maintain flow uniformity, the optimal manifold features the widest branch leading to outlet3 among the three terminal branches.

\begin{figure}[t]
	\centering
	\includegraphics[width=0.95\textwidth]{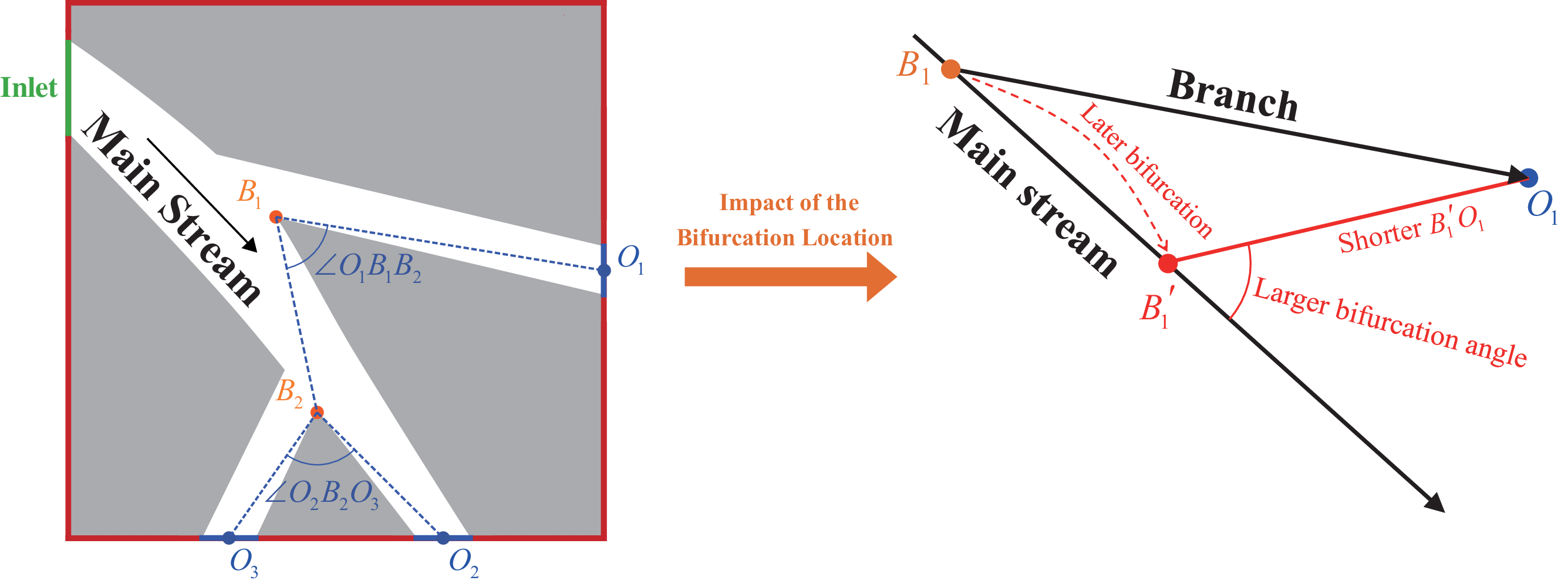}
	\caption{\label{fig:case1_topo}
		The general geometric configuration of the optimal manifold.
		$B_1$ and $B_2$ are the two bifurcation points, $O_1 \sim O_3$ are the central points of outlet boundaries. $\angle {O_1}{B_1}{B_2}$ and $\angle {O_2}{B_2}{O_3}$ are used to approximately characterize the magnitudes of the bifurcation angles.}
\end{figure}

\textbf{Wetted-area minimum}: 
As shown in \ref{fig:case1a}, the positions of the two bifurcation points, $B_1$ and $B_2$, largely determine the configuration of the optimized manifold. At Kn = 0.1, the two points are lower compared to those at Kn = 0.001 and Kn = 10, meaning that flow bifurcation occurs later in the streamwise direction. To analyze the underlying mechanism behind this shift in bifurcation location, we summarize the general geometric configuration of the optimal manifold in figure~\ref{fig:case1_topo}.
It can be seen that delaying the bifurcation points $B_1$ and $B_2$ along the main flow direction induces two competing effects:
\begin{quote}
(i) The bifurcation points are closer to the outlets, resulting in shorter branch channel lengths, which reduces wall friction losses and increasing MFR;\\
(ii) The bifurcation angles (characterized by $\angle {O_1}{B_1}{B_2}$ and $\angle {O_2}{B_2}{O_3}$) become larger, thereby increasing bend losses and decreasing MFR.
\end{quote}
Thus, the optimal bifurcation location represents a trade-off between wall friction loss and bend loss under a given Knudsen number: wall friction drives the optimal manifold toward shorter channel lengths and later bifurcations at the cost of larger bifurcation angles, whereas bend loss favors smaller bifurcation angles and earlier bifurcations at the expense of longer channels.

\begin{figure}[t]
 	\centering
 	\includegraphics[width=0.45\textwidth]{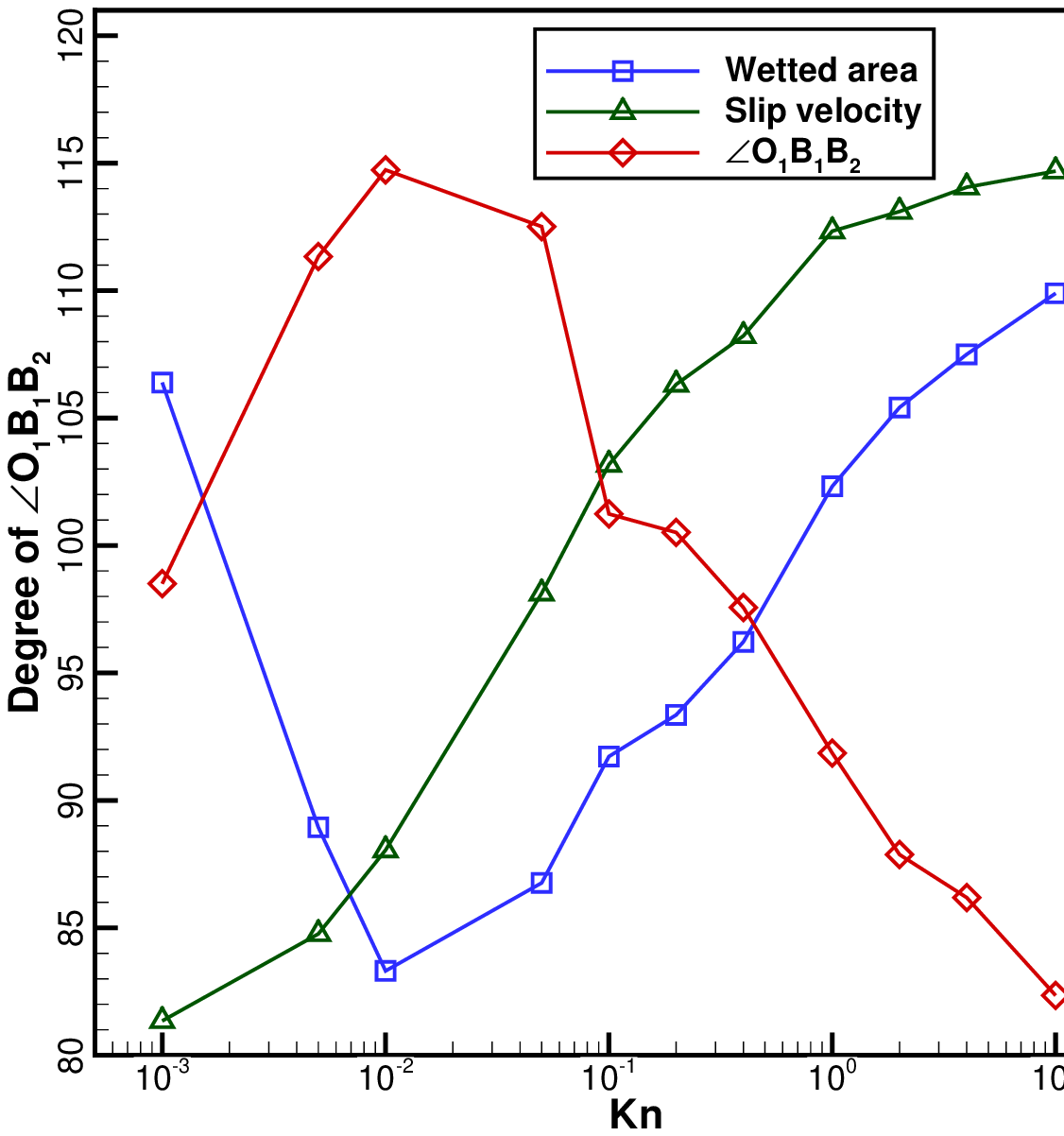}
 	\hspace{0.5cm}
	\includegraphics[width=0.45\textwidth]{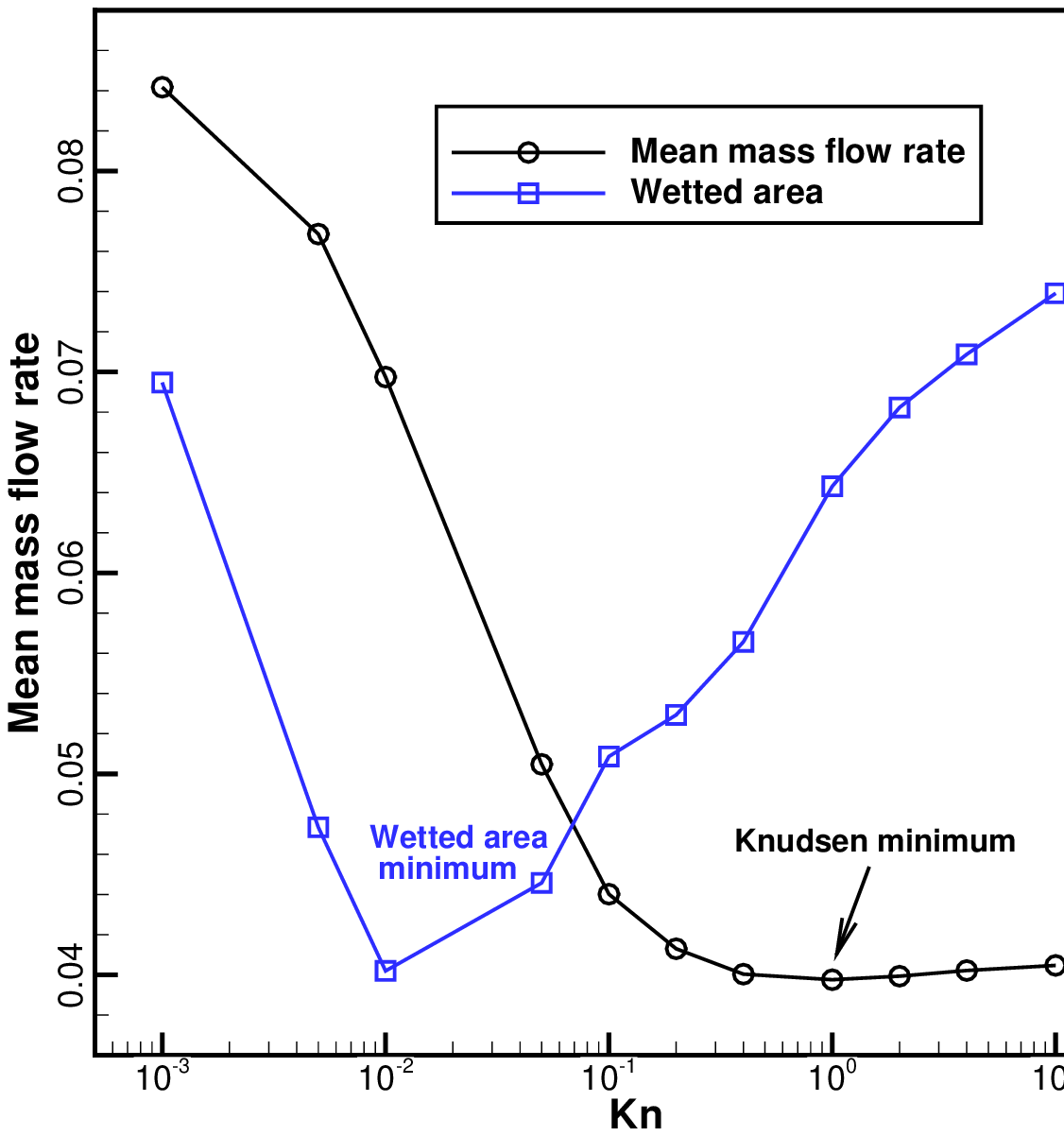}
 	\caption{\label{fig:case1_curve_warea}Optimization of a 2D intake manifold for MFR under flow uniformity constraint: the variations of the wetted area, the degree of $\angle {O_1}{B_1}{B_2}$, the slip velocity, and the mean MFR with the Knudsen number. The mean MFR is normalized by $\rho_{\rm in} a_\infty H$. The wetted area is obtained by measuring the perimeter of the flow passage (in the design domain the solid wall can be represented by the contour line $\theta=0.5$). The slip velocity is measured at point $S$ on the $x_1=4$ cross-section (shown in figure~\ref{fig:case1a}) and is normalized by the maximum velocity along that cross-section.}
 \end{figure}

To better illustrate this trade-off, we solved for the optimal manifold configurations from the slip to free-molecular regimes.
We quantify the channel length using the total wetted area within the design region (the contact area between the channel wall and the fluid, measured by the channel perimeter length in 2D), and use $\angle {O_1}{B_1}{B_2}$ to quantify the bifurcation angle. Their dependence on Kn is shown in figure~\ref{fig:case1_curve_warea}. Note that $\angle {O_2}{B_2}{O_3}$ exhibits a similar trend to $\angle {O_1}{B_1}{B_2}$, and is therefore omitted here for brevity. The figure reveals that the optimal wetted area and bifurcation angle exhibit completely opposite trends, which appears to reflect a geometric constraint inherent to the optimal manifold: one can only choose either a small wetted area with a large bifurcation angle, or vice versa. From a physical perspective, when wall friction is relatively stronger than bend loss, selecting a smaller wetted area is clearly more beneficial for the MFR. 

Therefore, the curve of optimal wetted area essentially reflects how the relative strength of wall friction changes with Kn. Figure~\ref{fig:case1_curve_warea} shows that the wetted area reaches a minimum in the range $0.01\lessapprox{\rm Kn}\lessapprox0.1$. This indicates that the relative impact of wall friction peaks in this regime and subsequently declines when Kn increases, accompanied by a rise in wetted area. Notably, this minimum occurs precisely within the slip regime, suggesting that velocity slip induced by rarefaction effects weakens the relative strength of wall friction. To investigate this hypothesis, we extract the slip velocity at point $S$ on the lower channel wall at the $x_1=4$ cross-section (see figure~\ref{fig:case1a}). The results are shown in figure~\ref{fig:case1_curve_warea}, and we can see a normalized slip velocity (relative to the maximum velocity along the cross-section) of approximately 0.2--0.3 near the wetted-area minimum, with a rapidly increasing trend---providing partial validation of our conjecture. In summary, the existence of this wetted-area minimum indicates that optimal designs near this Knudsen number should aim for a ``\emph{small wetted area}'', while at higher Knudsen numbers, the focus should gradually shift toward achieving a small bifurcation angle.

The curve of normalized mean MFR as a function of Kn is also presented in figure~\ref{fig:case1_curve_warea}, where the well-known Knudsen minimum~\cite{akhlaghi2023comprehensive} can be observed near Kn = 1. Notably, this MFR minimum does not coincide with the wetted-area minimum discussed above, but occurs at a distinctly different Knudsen number. This discrepancy likely arises because the wetted-area minimum is primarily associated with the onset of significant velocity slip at the wall, whereas the Knudsen minimum results from both wall velocity slip and the breakdown of the linear stress constitutive relation in the bulk region, thus the latter occurs at a higher Knudsen number (within the transition regime).



\subsection{Case 2: Optimization under reversed inlet--outlet conditions}\label{sec:intake_rvs}

\begin{figure}
\centering
\subfigure[\label{fig:case3a}Optimized configuration.]{
\includegraphics[width=0.33\textwidth]{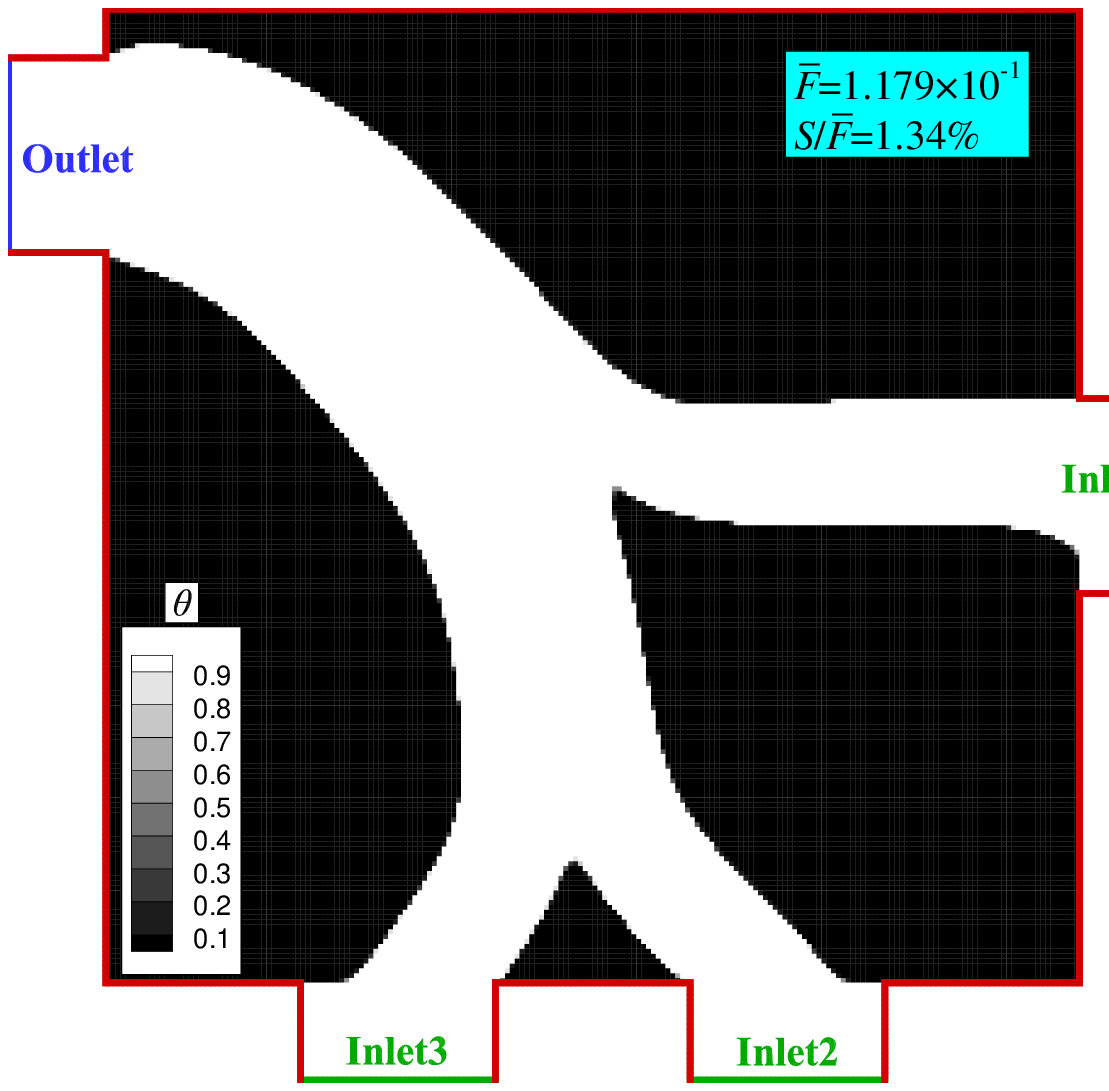}
\includegraphics[width=0.33\textwidth]{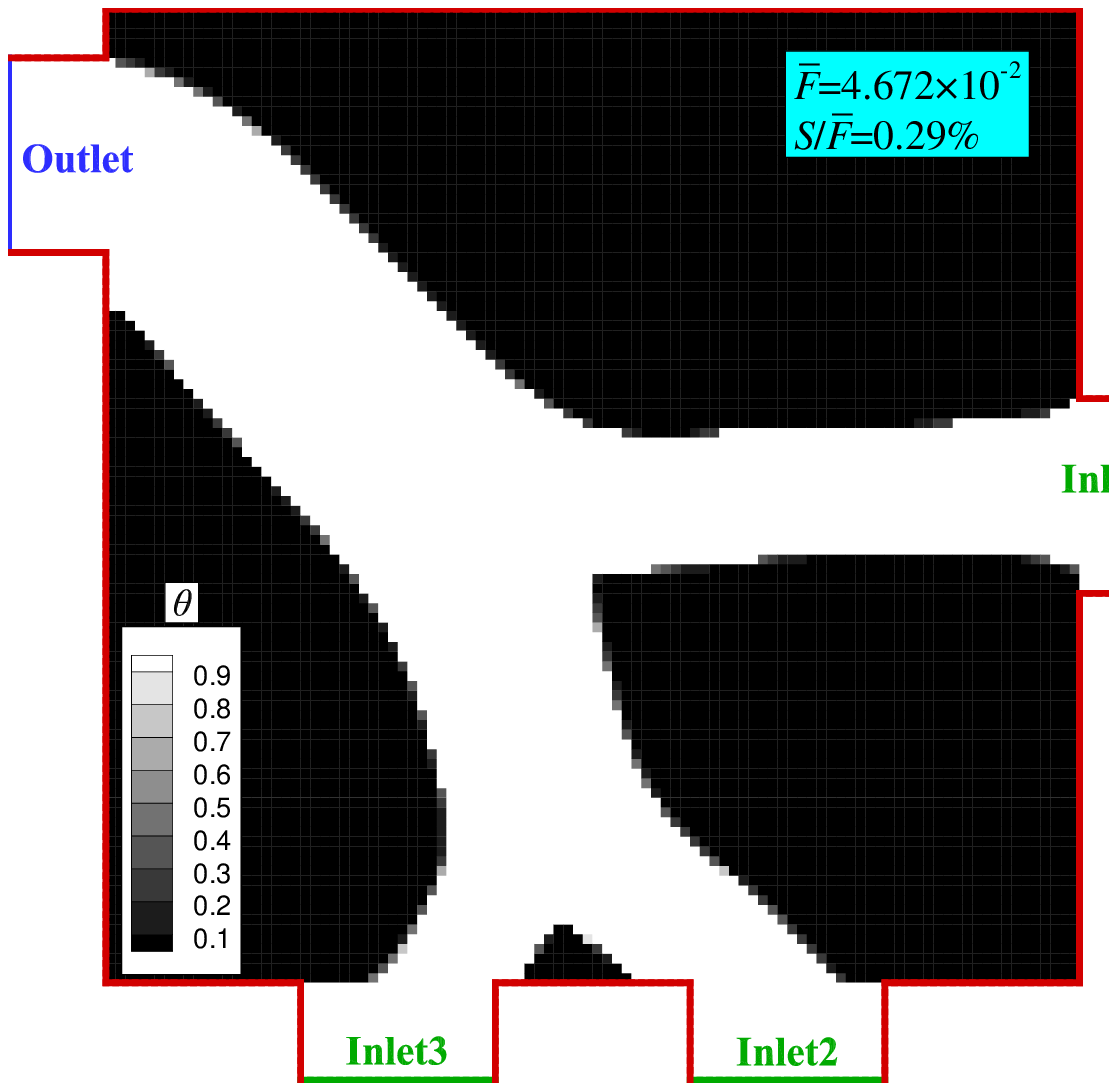}
\includegraphics[width=0.33\textwidth]{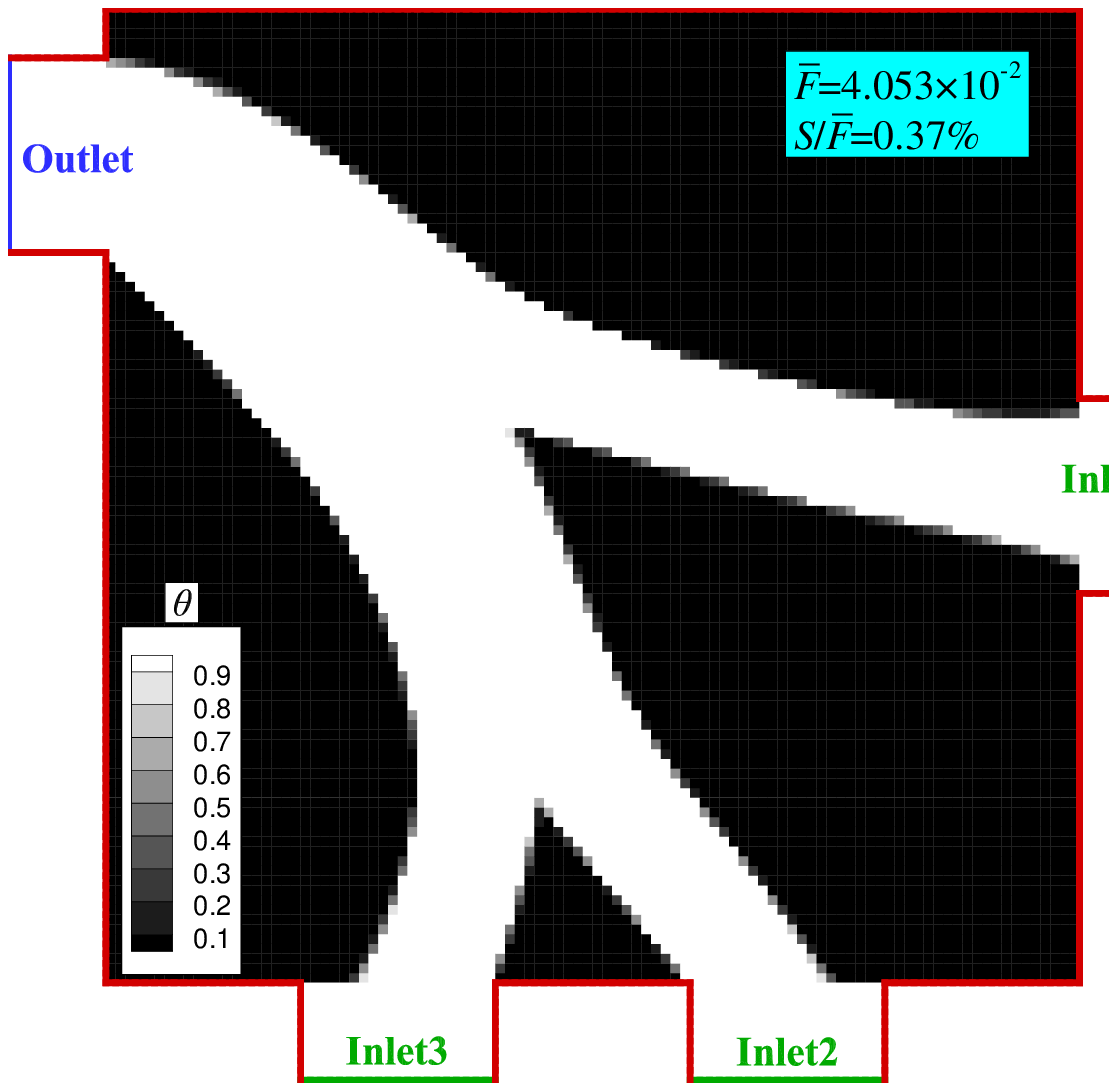}
}
\subfigure[\label{fig:case3b}Optimal configuration changes represented by $\Delta \theta$, relative to Case 1 (Section \ref{sec:intake}). $\Delta \bar F/{\bar F_{{\rm{case1}}}}$ is the relative flow rate increment compared to the results of Case 1.]{
\includegraphics[width=0.33\textwidth]{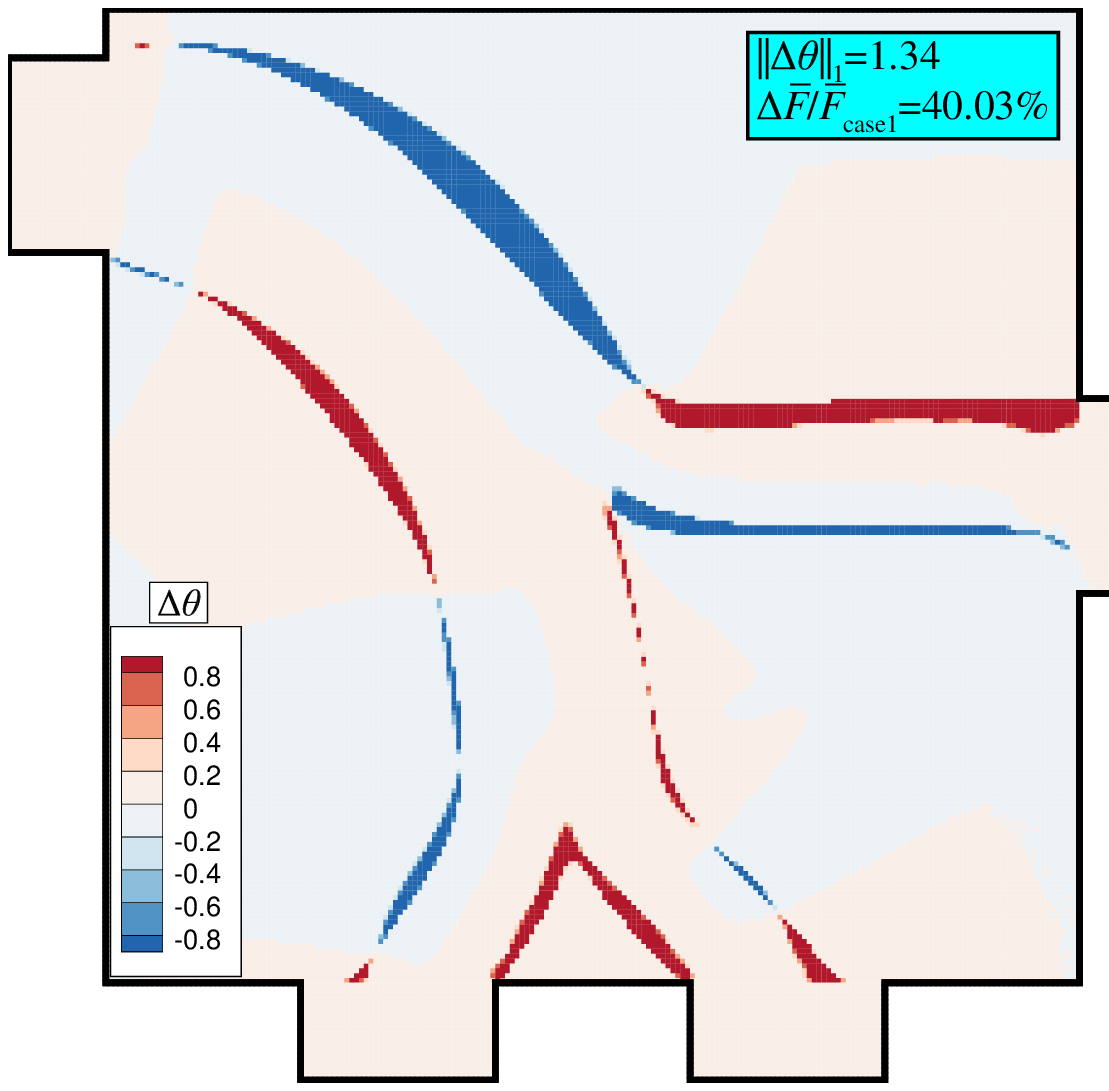}
\includegraphics[width=0.33\textwidth]{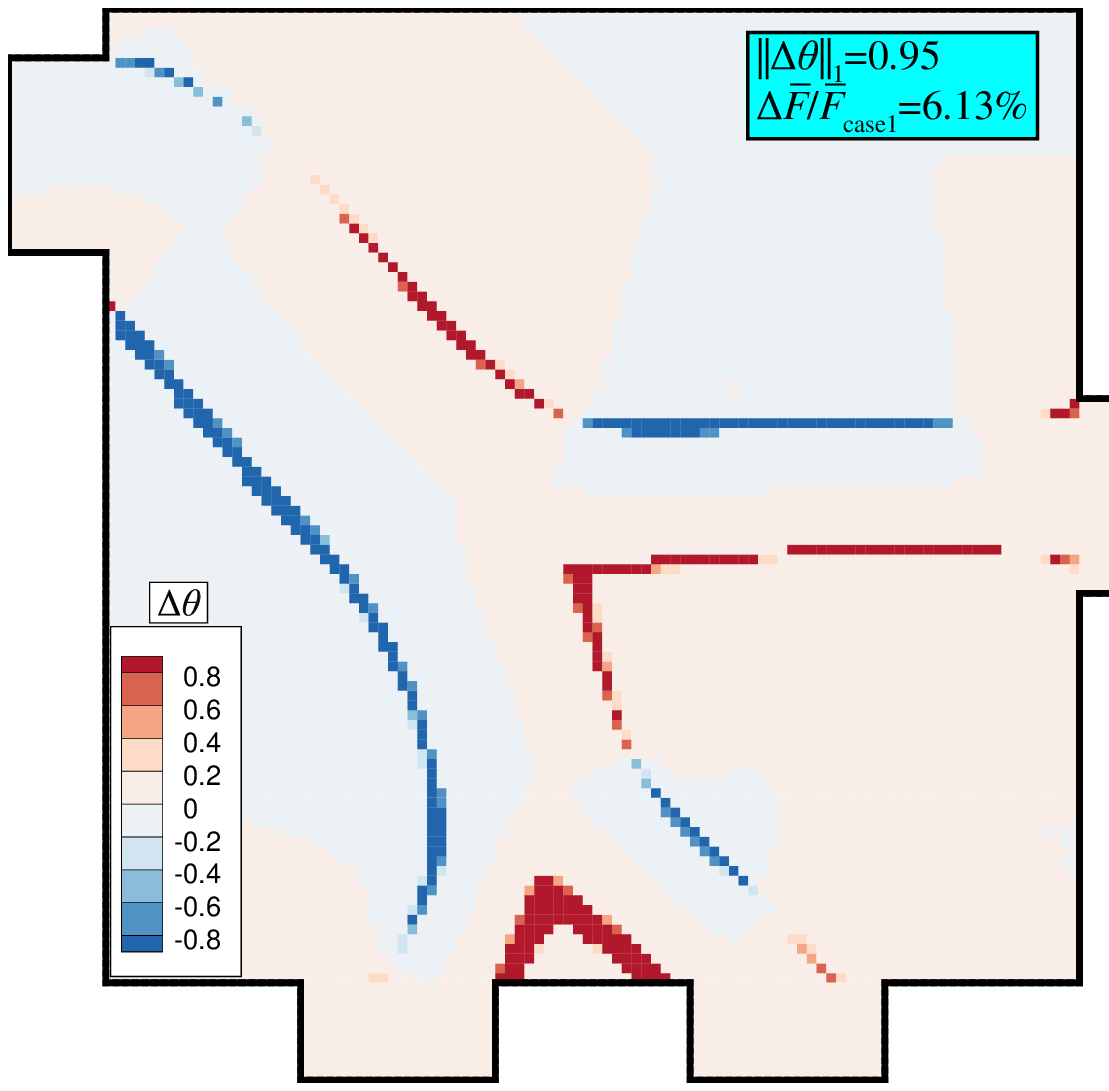}
\includegraphics[width=0.33\textwidth]{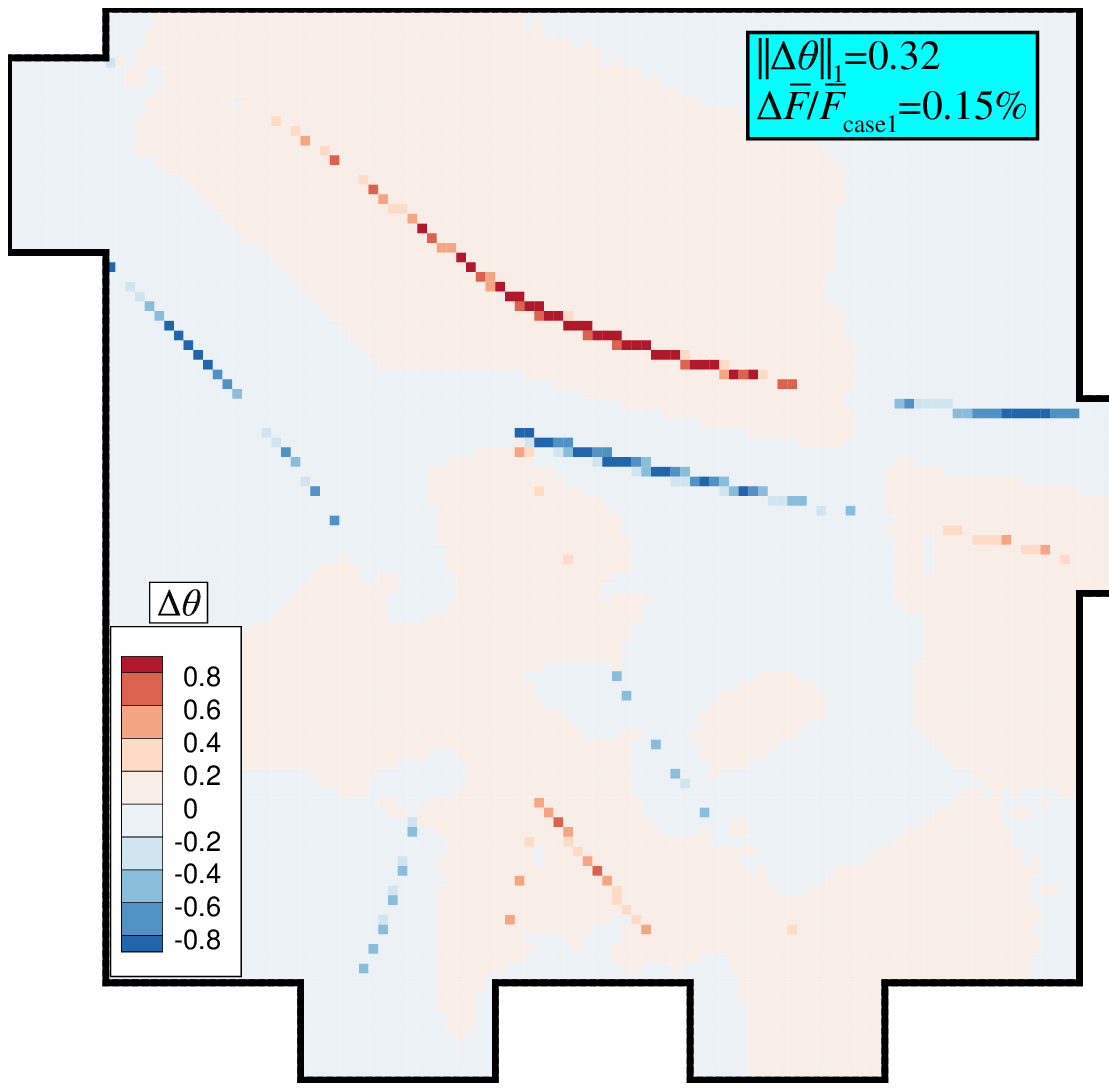}
}
\subfigure[Distribution of Mach number]{
\includegraphics[width=0.33\textwidth]{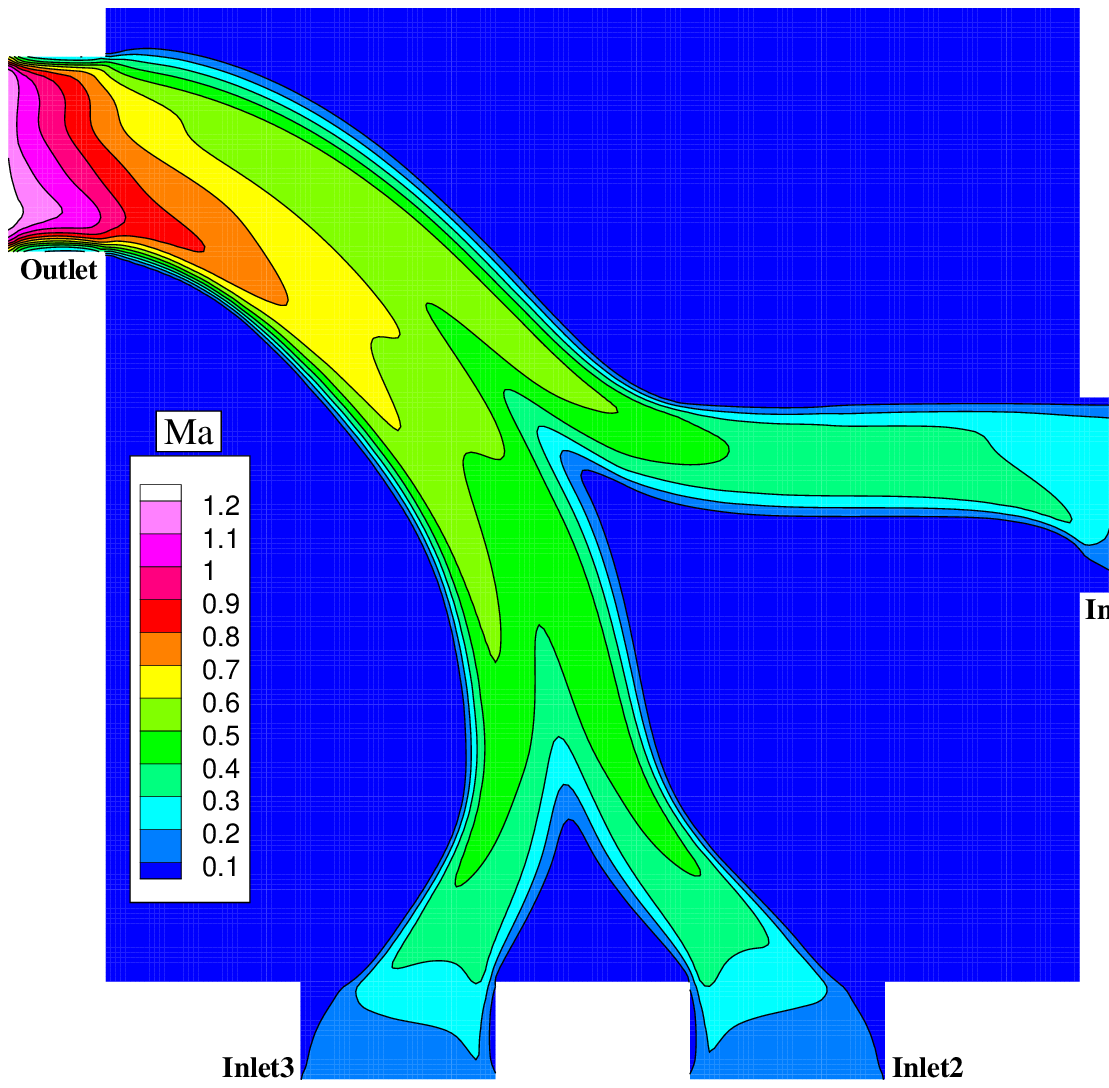}
\includegraphics[width=0.33\textwidth]{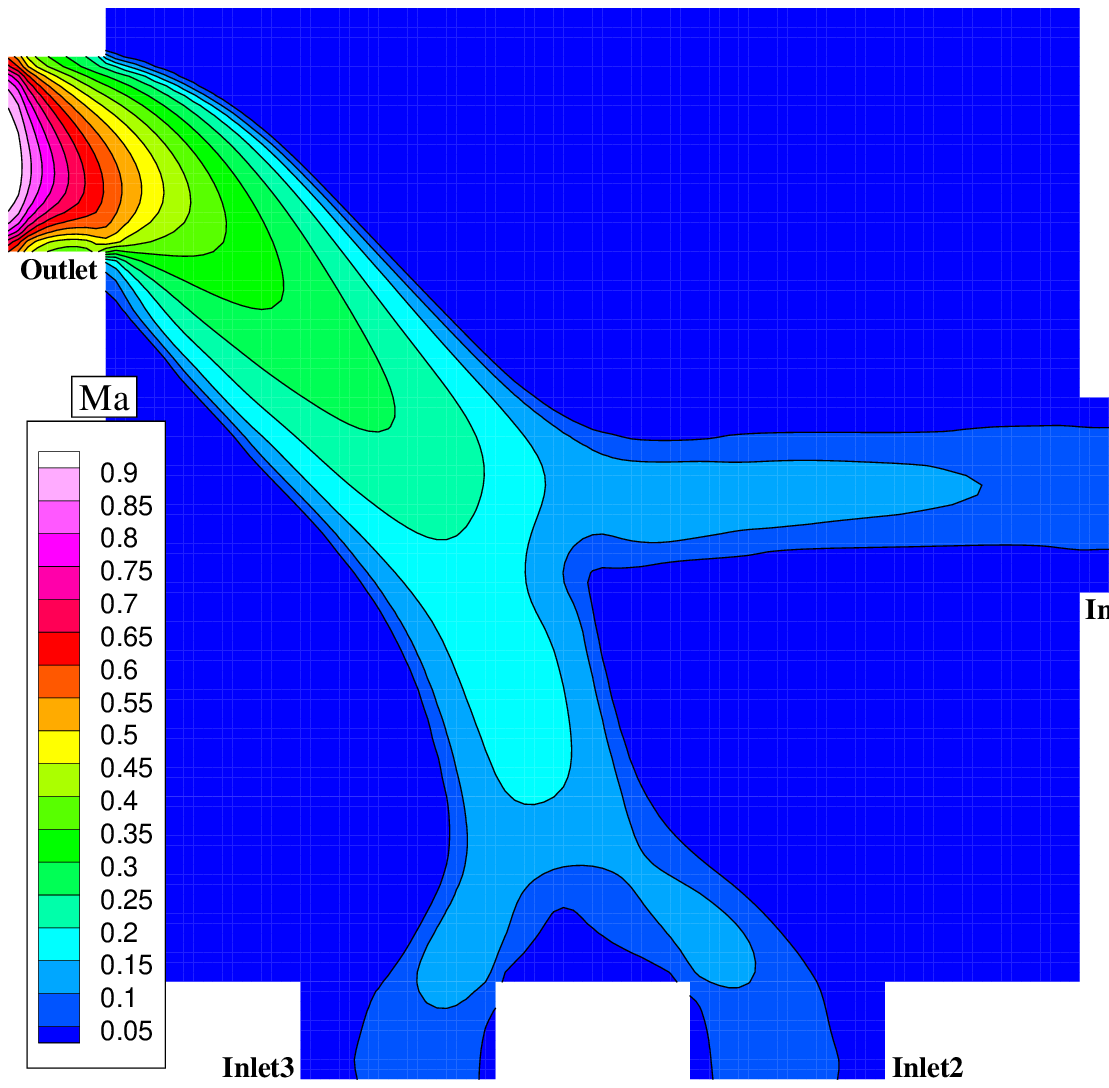}
\includegraphics[width=0.33\textwidth]{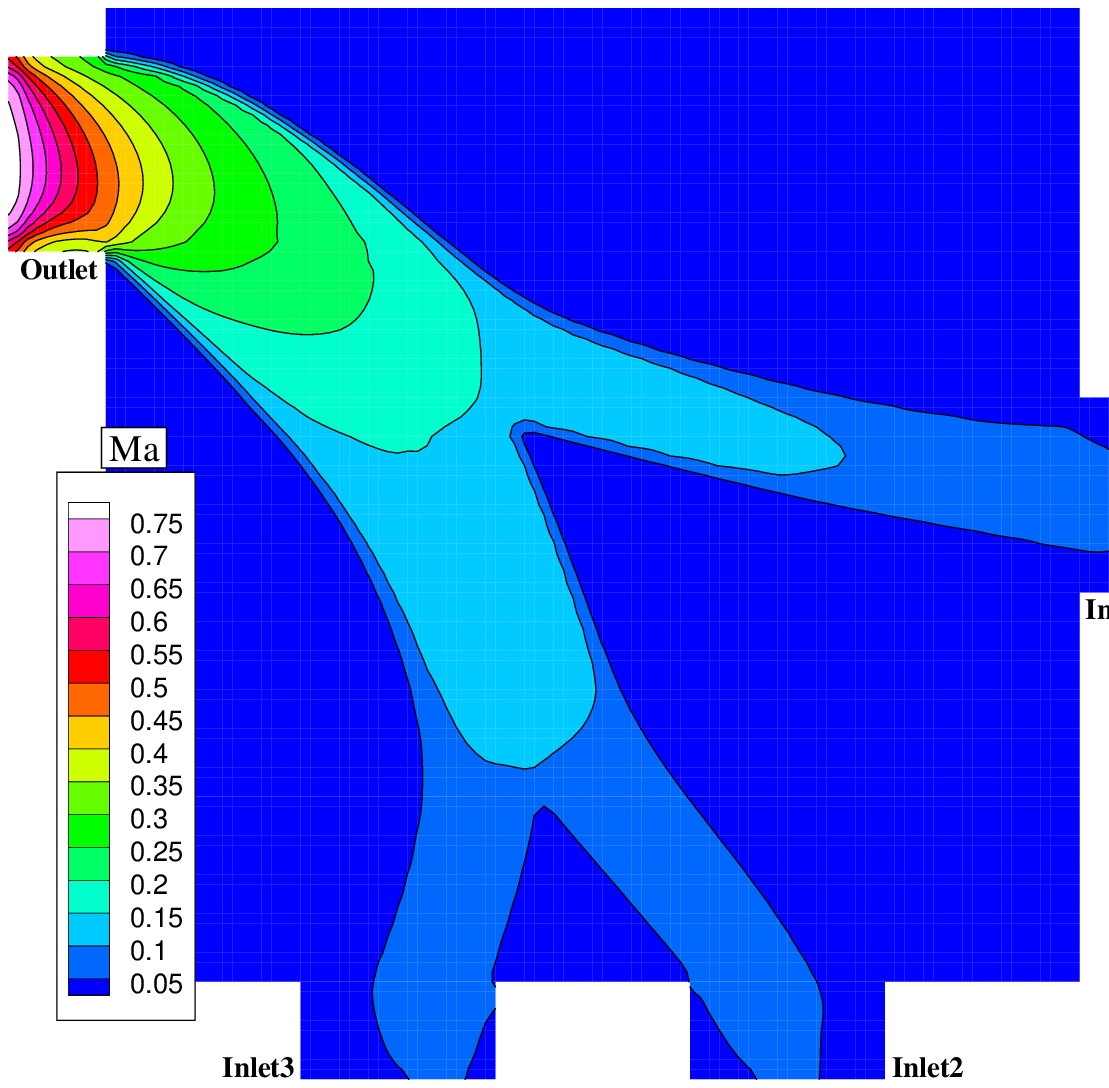}
}
\caption{\label{fig:case3}Optimization of the 2D manifold under reversed inlet--outlet conditions. From left to right are the results for ${\rm Kn}=0.001,0.1,$ and 10, respectively.
}
\end{figure}

We perform optimizations by exchanging the inlet--outlet conditions of Case 1 (Section \ref{sec:intake}), and investigate how optimal solutions would change under different degrees of gas rarefaction. Namely now there are three inlets: inlet1$\sim$inlet3 (originally outlet1$\sim$outlet3), one outlet (originally the inlet), and the objective is to increase the MFR meanwhile maintaining uniformity of the flow rate among the three inlets. 

The optimized results are shown in figure~\ref{fig:case3}. Regarding the variation of key geometric features of the optimal manifold (such as bifurcation point positions and wetted area) with the Knudsen number, the trends are essentially the same as those in Case 1 and thus will not be repeated here. Instead, we focus primarily on the changes in the optimal manifold configuration relative to the Case 1. 
The difference in the optimal material density obtained between Case 2 and Case 1, as well as the difference in MFR, are examined. These two differences are calculated as
\begin{equation}
\begin{aligned}
 \Delta \theta  = {\theta _{{\rm{case2}}}} - {\theta _{{\rm{case1}}}}, \quad
\frac{\Delta \bar F}{{\bar F}_{{\rm{case1}}}} = \frac{{{\bar F}_{{\rm{case2}}}} - {{\bar F}_{{\rm{case1}}}}} {{\bar F}_{{\rm{case1}}}}.
\end{aligned}
\end{equation}
It can be seen from figure~\ref{fig:case3b} that, as the Knudsen number increases, both the geometric changes in the optimal manifold and the corresponding variations in MFR diminish: for Kn~= 0.001, 0.1, and 10, $\Delta \bar F/{{\bar F}_{{\rm{case1}}}}$ are 40.03\%, 6.13\%, and 0.15\%, respectively.
 
In other words, under high Knudsen numbers, exchanging inlet and outlet boundary conditions produces an optimal configuration and MFR that are almost unchanged from the pre-exchange state, reflecting a type of inlet–outlet reciprocity. This inlet–outlet reciprocity can be well explained by the following theoretical consideration:
\begin{quote}
In the free-molecular limit (${\rm{Kn}} \to \infty $), the gas flow field is governed by the collisionless Boltzmann equation, which is a homogeneous linear equation and thus satisfies the superposition principle of solutions. This implies that for any given manifold configuration, the flow with inlet pressure $p_1$ and outlet pressure $p_2$, when superimposed with the flow with inlet pressure $p_2$ and outlet pressure $p_1$, should equal the flow under uniform pressure $p_1+p_2$ at both ends, where the MFR is clearly zero. Consequently, the magnitude of MFR in each channel of the manifold remains unchanged upon exchanging the inlet and outlet pressures, and therefore flow-rate-based optimizations (e.g. maximizing total MFR and/or controlling its distribution among different channels) should yield identical results.
\end{quote}
We have verified that, based on the case of this section, altering the positions of the inlet/outlet or changing the pressure ratio across them does not affect the validity of the above  inlet–outlet reciprocity under free-molecular flow regime. Furthermore, the above analysis leads to an interesting property: \emph{in the free-molecular limit, flow-rate-based optimization of a channel system yields identical optimal designs regardless of the inlet--outlet pressure ratio.}
These findings also suggest that it is fundamentally impossible to design a passive directional flow device (such as a ``Tesla valve'') in free-molecular flow.
Additionally, similar property has in fact been reflected in previous studies on the gas flow diode effect through a single tapered channel, where the ratio of MFRs after exchanging the inlet--outlet pressures is unity in the free-molecular limit~\cite{szalmas2015analysis,graur2016physical,hemadri2018liquid}.

Another point worth mentioning is that, under the small Knudsen number of ${\rm Kn}=0.001$, compressibility has a significant impact on the results. As shown in figure~\ref{fig:case3}, after exchanging the inlet-outlet conditions, the flow field inside the manifold is almost entirely subsonic. Only at the outlet does the flow reach supersonic, resulting in only one nozzle structure forming near the outlet. This also leads to a substantial \textasciitilde40\% increase in MFR compared to Case 1, as the flow no longer undergoes repeated shock wave formation within the manifold, thereby saving a lot of energy losses. Such a pronounced effect of compressibility is not found under the rarefied conditions of ${\rm Kn}=0.1$ and 10.
This observation is consistent with the findings in Ref.~\cite{hemadri2018liquid}, where the authors reported that a single tapered channel exhibits high diodicity under low Knudsen numbers and high Mach numbers due to compressibility effects.


\subsection{Case 3: Optimization without flow uniformity constraint}\label{sec:intake_np}

\begin{figure}
	\centering
	\subfigure[\label{fig:case2a}Optimized configuration without uniformity constraint]{
		\includegraphics[width=0.33\textwidth]{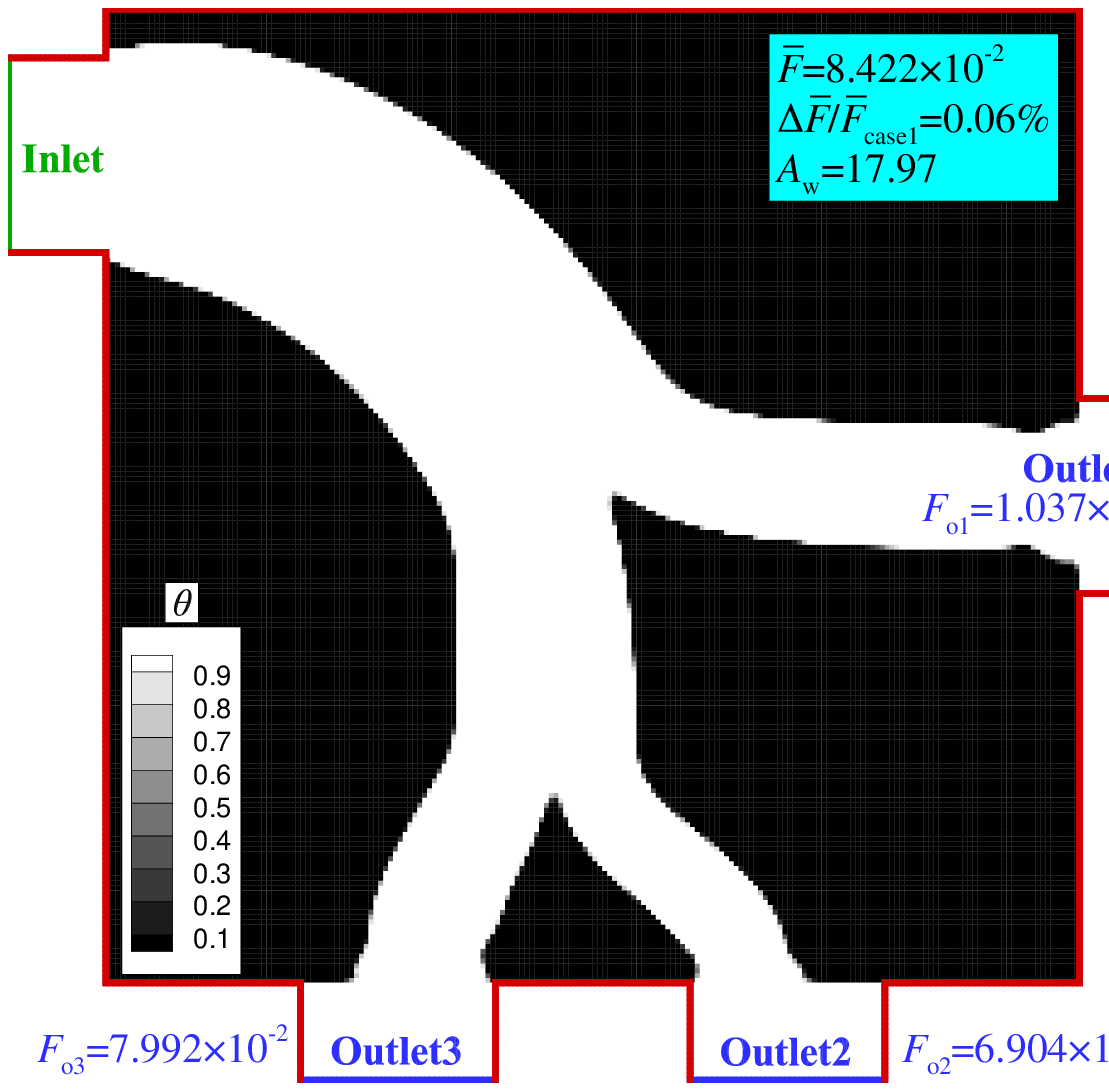}
		\includegraphics[width=0.33\textwidth]{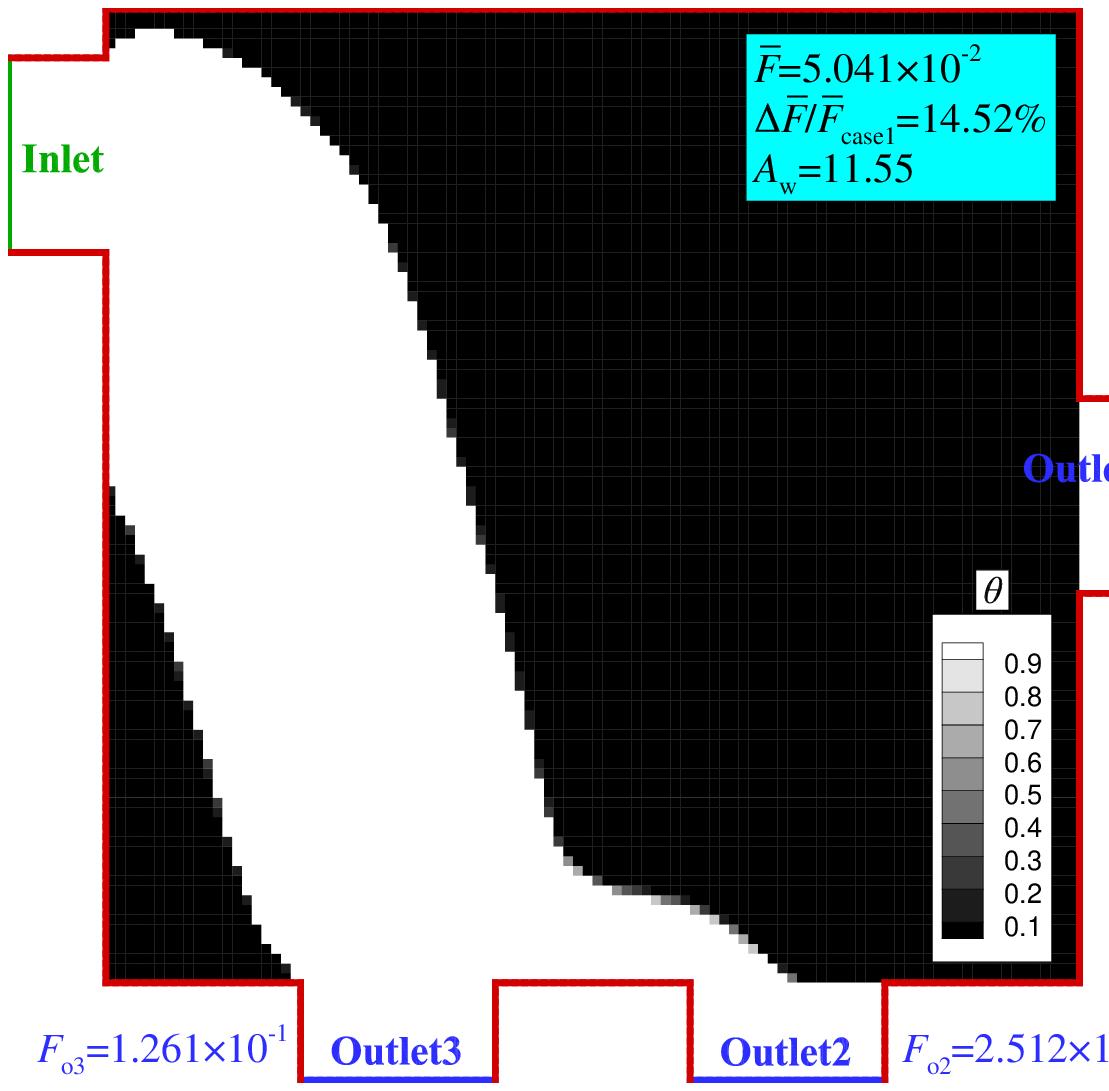}
		\includegraphics[width=0.33\textwidth]{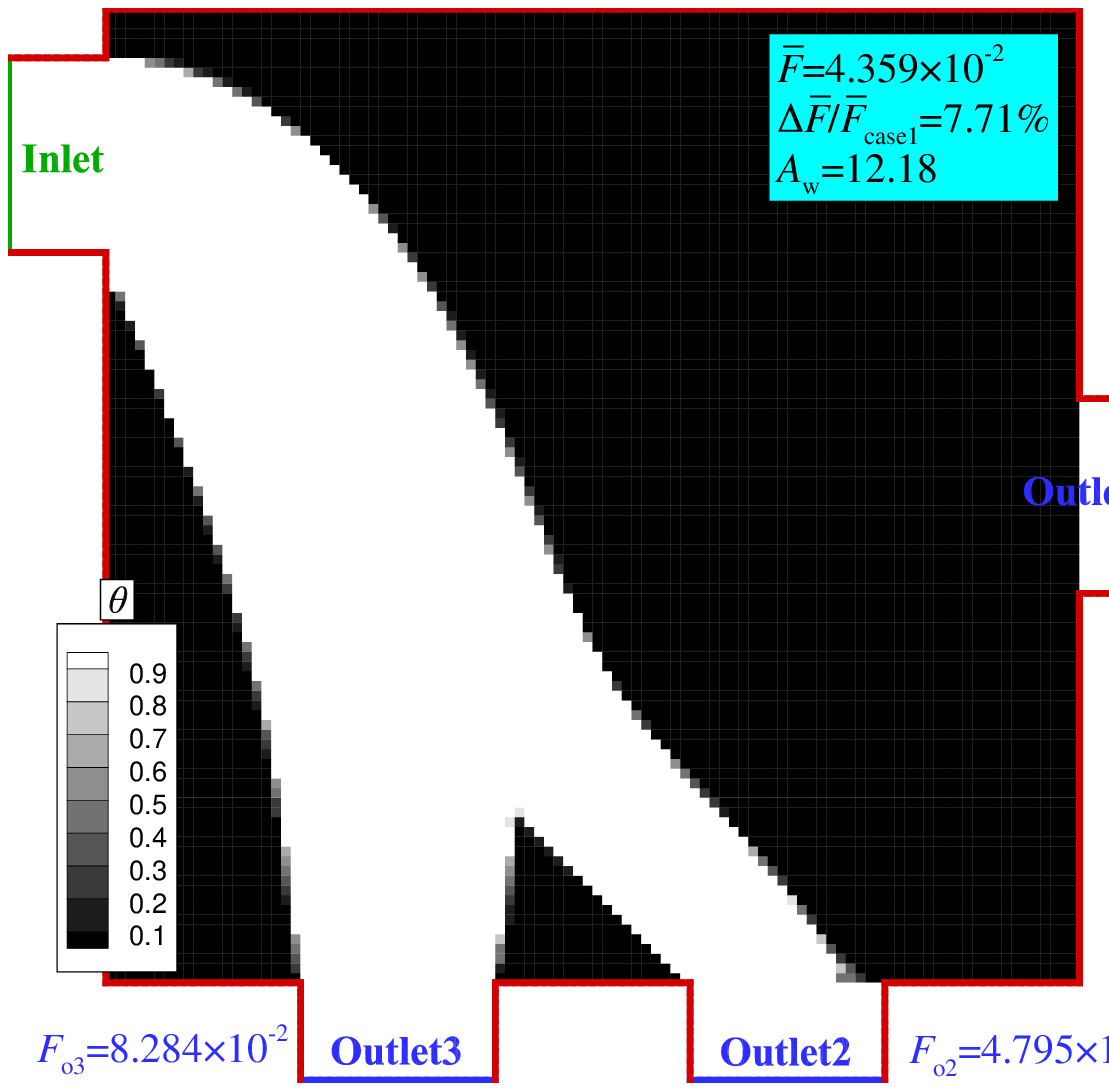}
	}
	\subfigure[Distribution of Mach number]{
		\includegraphics[width=0.33\textwidth]{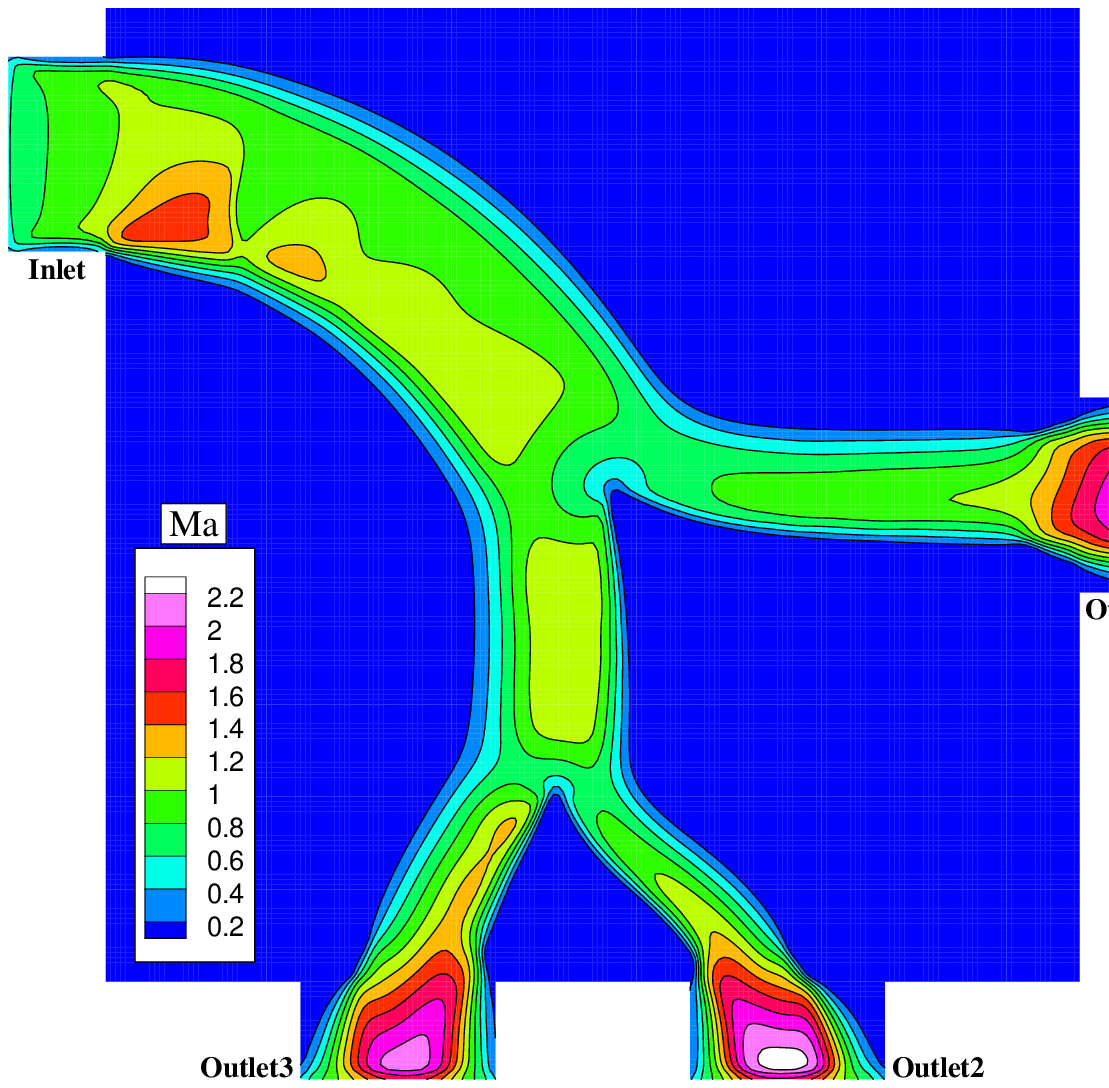}
		\includegraphics[width=0.33\textwidth]{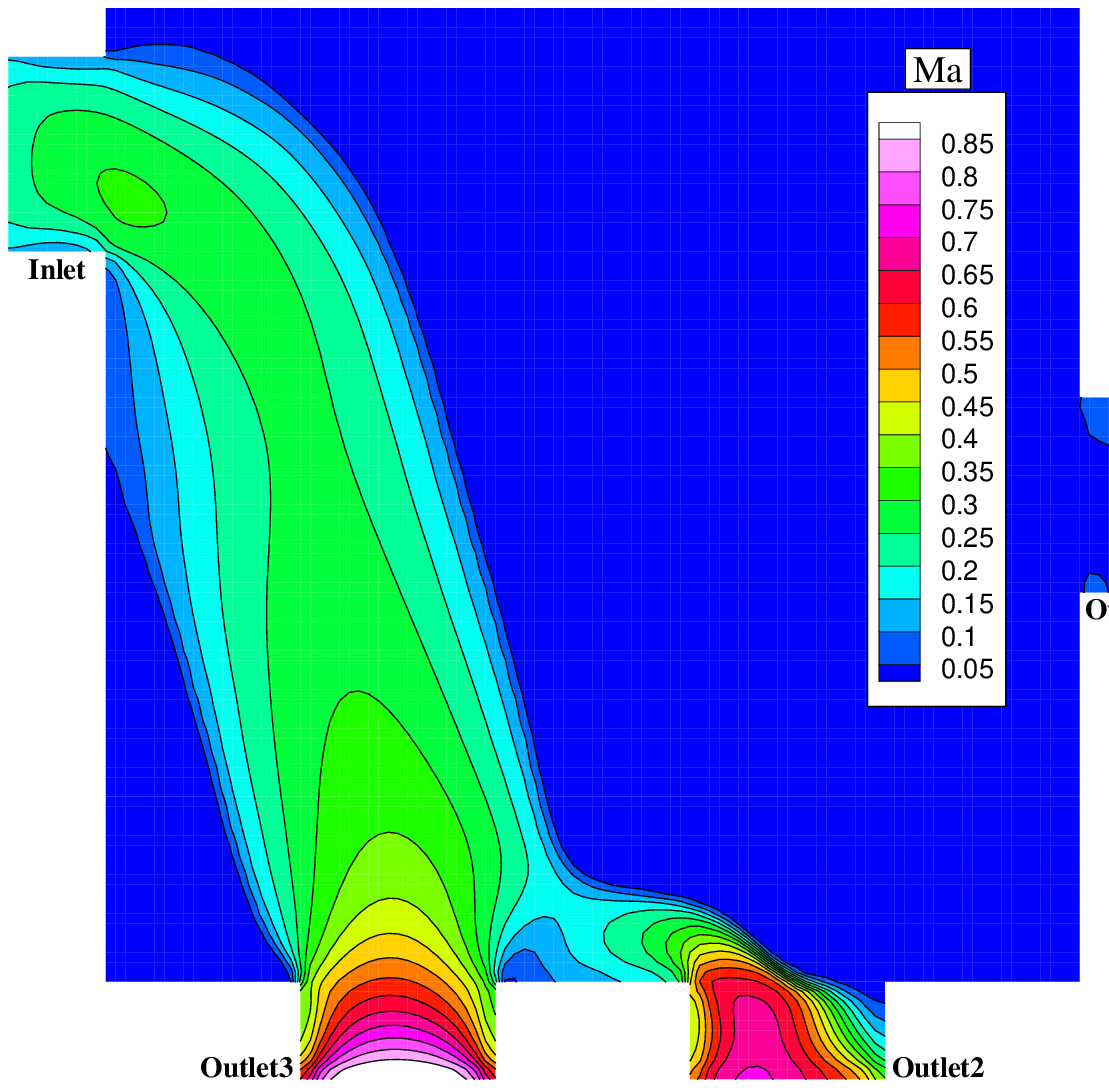}
		\includegraphics[width=0.33\textwidth]{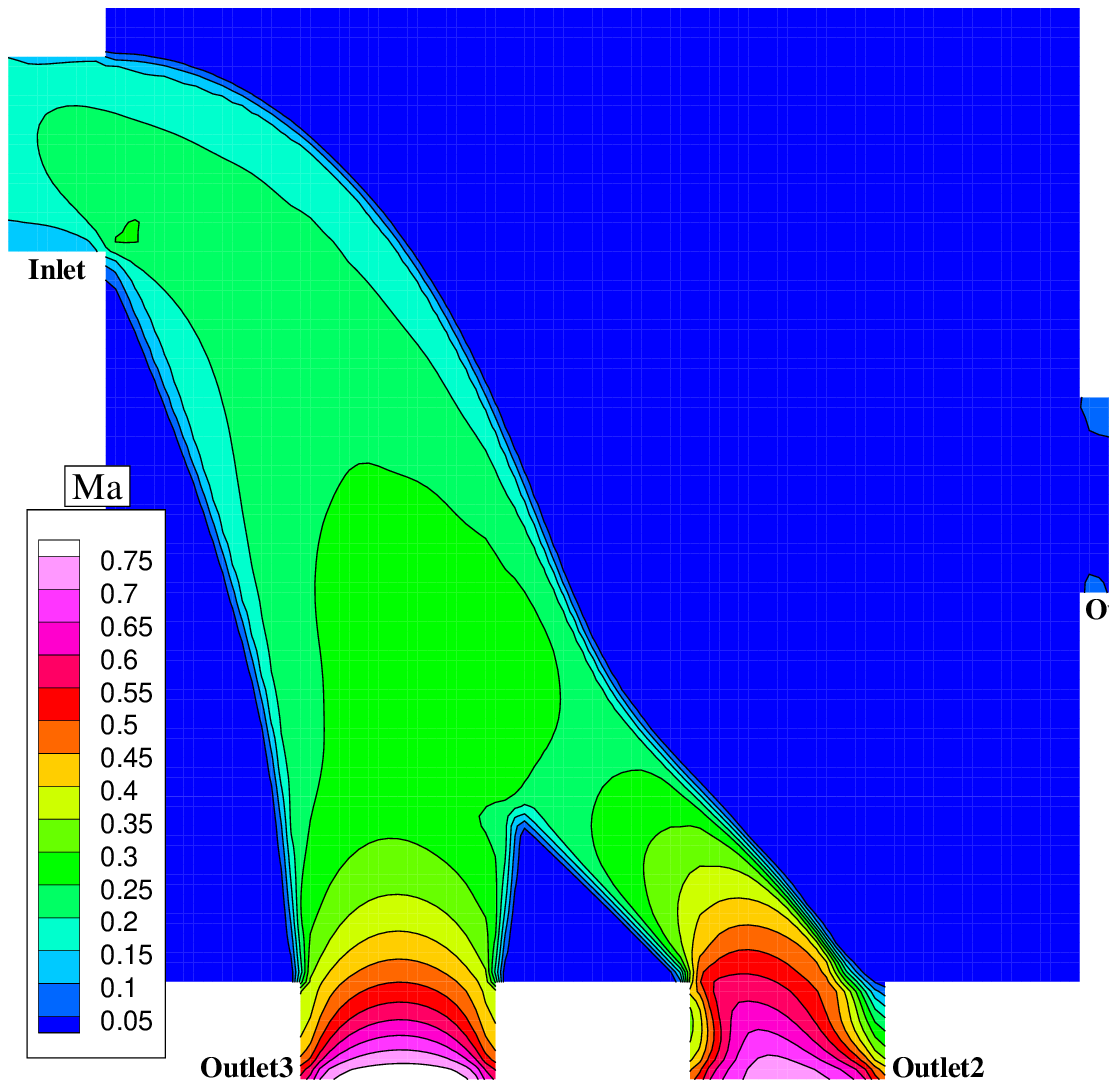}
	}
	\subfigure[Streamlines and pressure contours\label{fig:case2c}]{
		\includegraphics[width=0.33\textwidth]{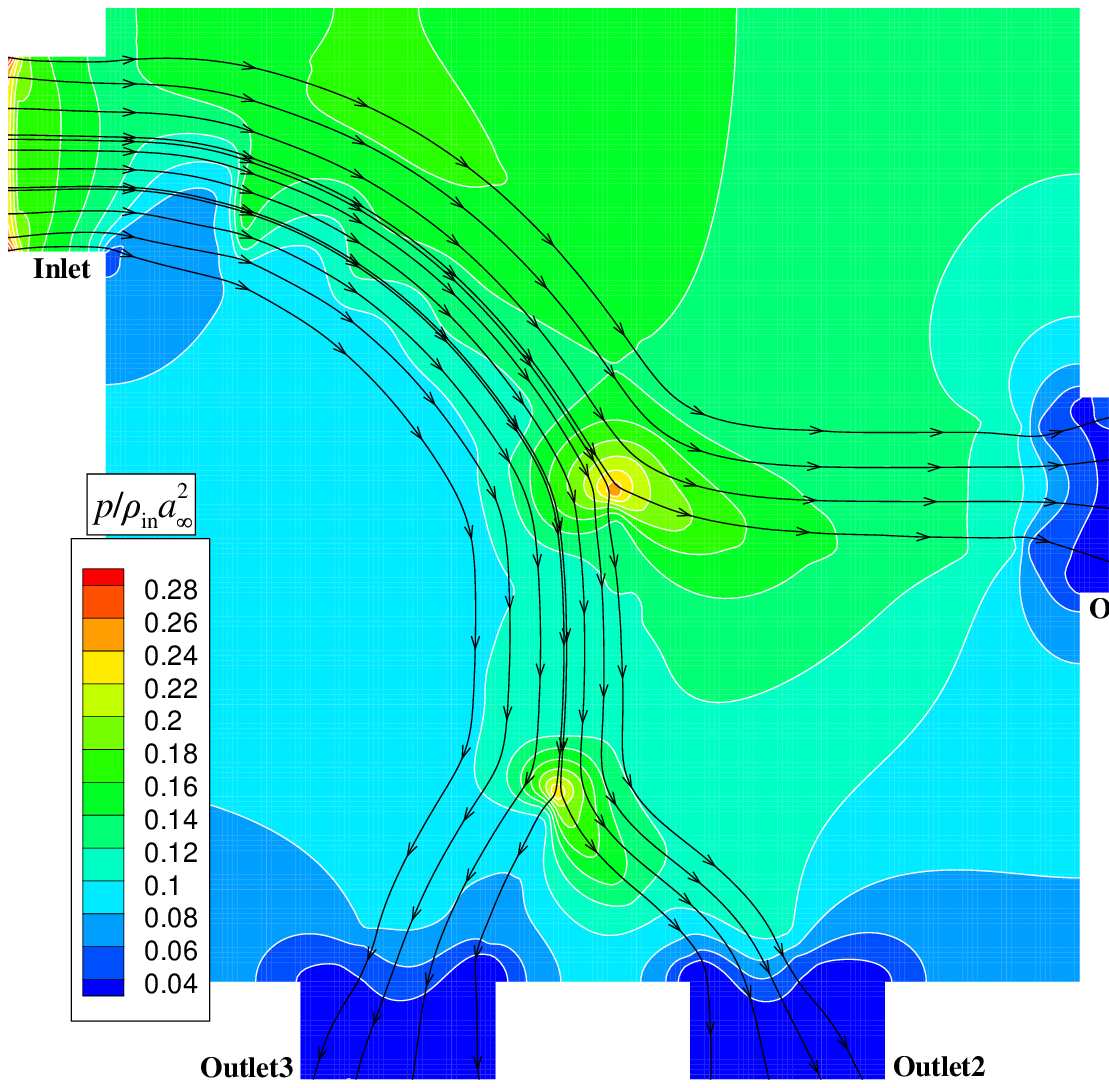}
		\includegraphics[width=0.33\textwidth]{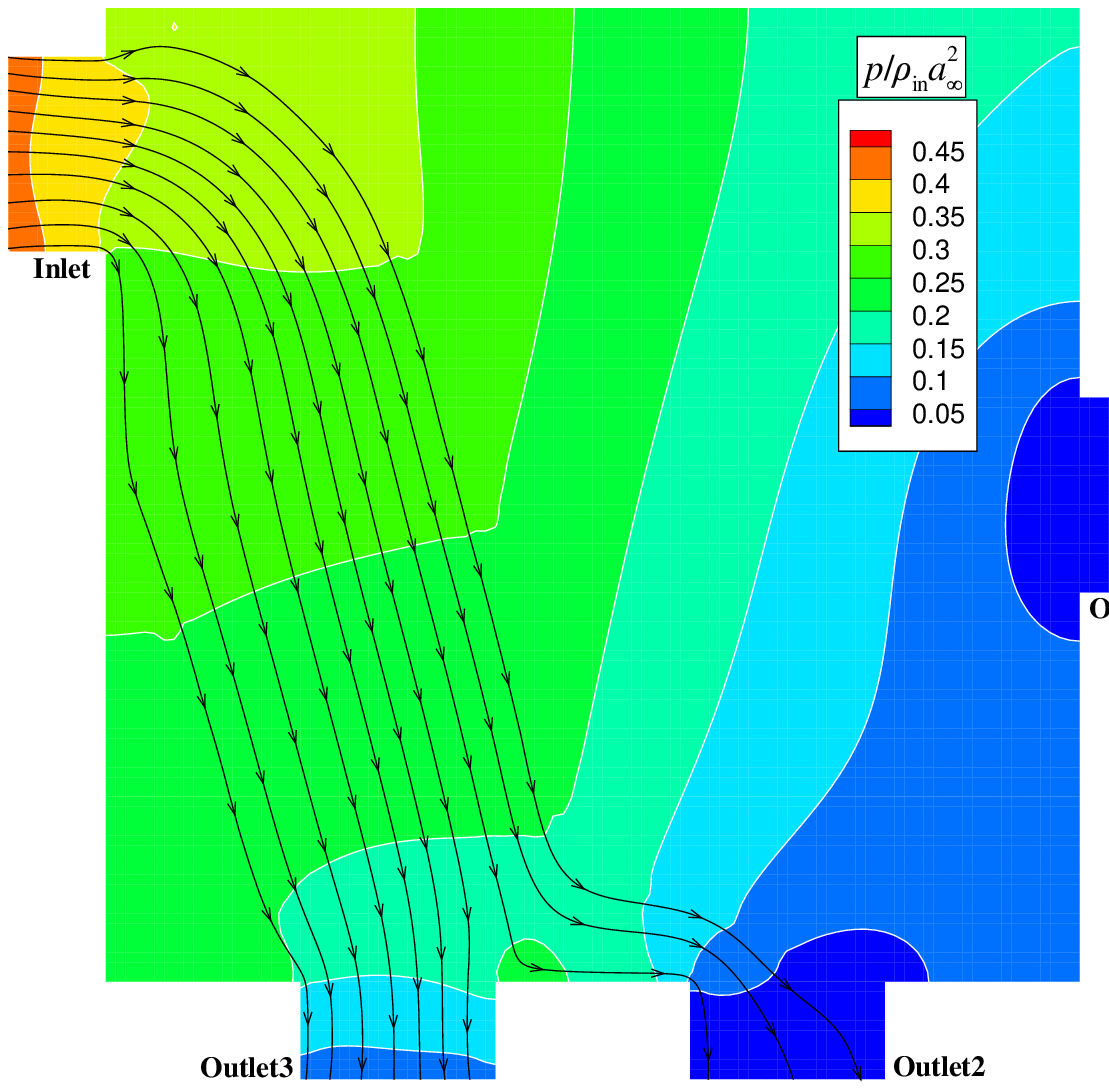}
		\includegraphics[width=0.33\textwidth]{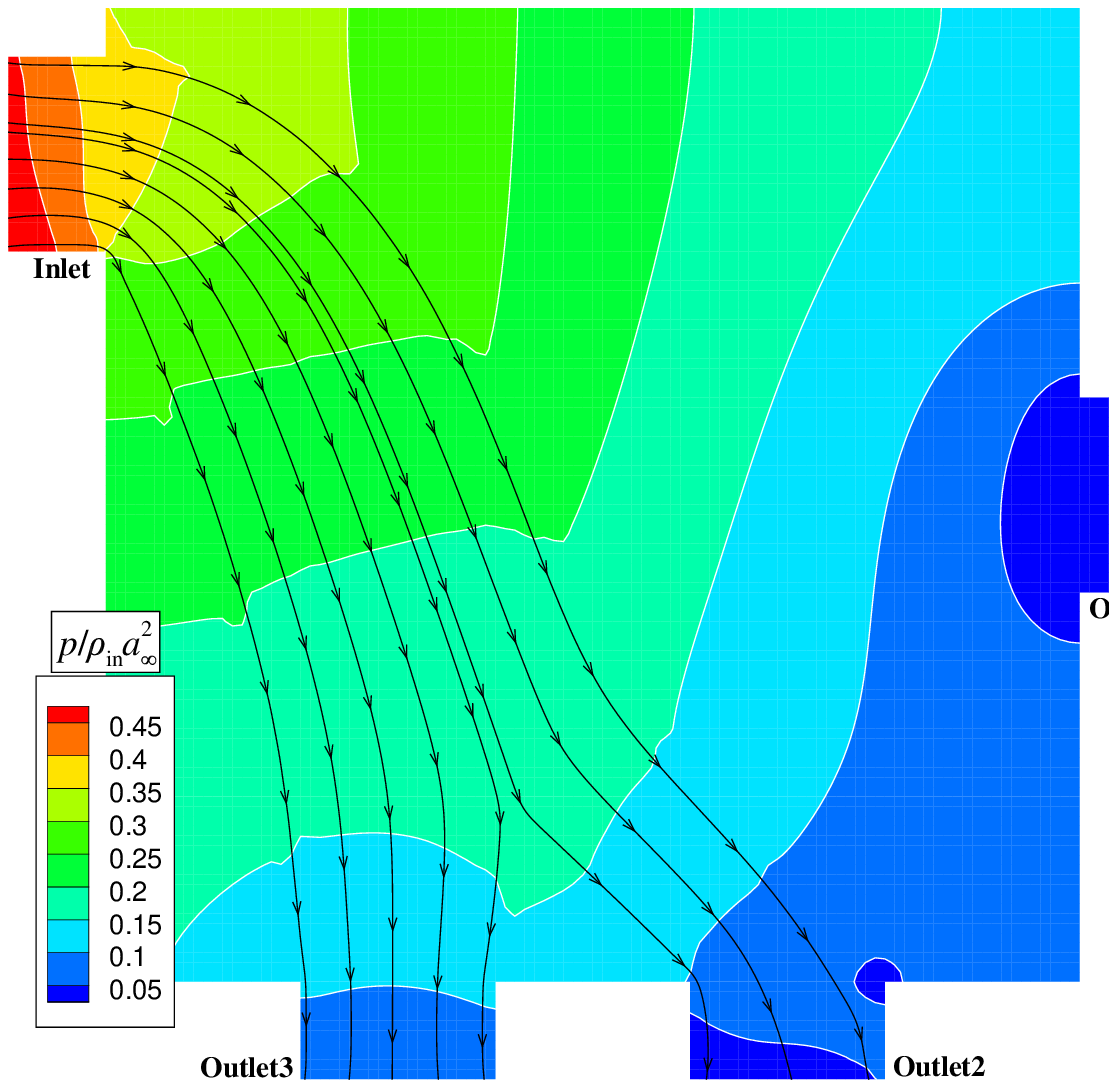}
	}
	\caption{\label{fig:case2}Optimization of the 2D manifold without flow uniformity constraint. From left to right are the results for ${\rm Kn}=0.001,0.1,$ and 10, respectively. Note that the contour levels for different cases are intentionally set identical to their counterparts in figure~\ref{fig:case1} to facilitate direct comparison.}
\end{figure}

To investigate the impact of the flow uniformity constraint, topology optimizations are performed by setting the penalty factor $\epsilon=0$ in equation~\eqref{eq:obj}. 
All other settings are maintained identical to those in Case 1. 

Note that in the case of this section, for ease of comparison with the results of Case 1, the mean MFR $\bar F$ is still computed as $1/3$ of the total MFR (even if some outlets are not connected). Accordingly, the relative total MFR increment is calculated as:
\begin{equation}
\frac{\Delta \bar F}{{\bar F}_{{\rm{case1}}}} = \frac {{{\bar F}_{{\rm{case3}}}} - {{\bar F}_{{\rm{case1}}}}} {{\bar F}_{{\rm{case1}}}}.
\end{equation}

The optimized manifolds are shown in figure~\ref{fig:case2}. Without the flow uniformity constraint, the manifolds optimized under the three Knudsen numbers exhibit entirely different topologies, each with a distinct number of channel branches. For Kn = 0.1 and 10, only outlet2 and outlet3 are connected. The MFRs for each case and each branch are summarized in table \ref{tab:case_np}.

\begin{table}[h]
	\centering
	\caption{\label{tab:case_np}Optimization of the 2D manifold without flow uniformity constraint: the optimized MFR, and the total MFR increase relative to Case 1. The MFR is normalized by $\rho_{\rm in} a_\infty H$.}
	\begin{tabular}{cccccc}
		\hline
		Kn    & $F_{\rm o1}$       & $F_{\rm o2}$       & $F_{\rm o3}$       & $\bar F$  & ${\Delta \bar F}/{{\bar F}_{{\rm{case1}}}}$ 
		 \\ \hline
		0.001 & $1.037\times 10^{-1}$ & $6.904\times 10^{-2}$ & $7.992\times 10^{-2}$ & $8.422\times 10^{-2}$    & 0.06\%                                                  \\
		0.1   & - & $2.512\times 10^{-2}$ & $1.261\times 10^{-1}$ & $5.041\times 10^{-2}$    & 14.52\%                                                  \\
		10    & - & $4.795\times 10^{-2}$ & $8.284\times 10^{-2}$ & $4.359\times 10^{-2}$    & 7.71\%                                                  \\ \hline
	\end{tabular}
\end{table}


These topological variation can likewise be interpreted in terms of wall friction and wetted-area effects. The primary advantage of reducing the number of branches lies in the decreased wetted area and the consequent widening of channels (due to the volume constraint), which together reduces wall-friction losses. Accordingly, at Kn = 0.001, where wall friction has a relatively minor influence, both the optimal manifold configuration and the total MFR remain almost unchanged from those obtained under the flow uniformity constraint. 

In contrast, at higher Knudsen numbers (Kn = 0.1 and 10), where the flow is more sensitive to wall friction, the optimal manifolds reduce the number of channel branches to minimize the wetted area, leading to pronounced configuration changes. In particular, at Kn = 0.1—which is close to the wetted-area minimum discussed in Section~\ref{sec:intake}—the optimal manifold contains primary a single flow channel, where the inlet--outlet distance is smallest among possible solutions, achieving the smallest wetted area among the three cases and a marked increase in total MFR. This again highlights the ``\emph{small wetted area}” design principle. 
Of course, this comes at the cost of a larger bifurcation angle: as seen from the streamline of figure \ref{fig:case2c}, at the bottom of the manifold, the bifurcation point where the flow splits toward outlet2 and outlet3 exhibits a bifurcation angle of $180^\circ$.

For the optimal manifold at Kn = 10 , although it also connects only outlet2 and outlet3---just like the case at Kn = 0.1---it develops a distinct bifurcation structure upstream of the outlets, featuring a much smaller bifurcation angle. Naturally, this comes at the cost of a larger wetted area. This comparison is fully consistent with the analysis in Case 1 regarding the trade-off between wetted area and bifurcation angle.

By the way, from the above analysis, it is concluded that the effect of the flow uniformity constraint on the total MFR is primarily governed by wall friction, whose influence decreases in the order of 0.1, 10, 0.001. This also explains why, in Case 1, a large flow-uniformity penalty factor $\epsilon=300$ is used for ${\rm Kn}=0.1$ and 10, while a small value $\epsilon=10$ is used for ${\rm Kn}=0.001$.

\section{Conclusions}\label{sec:conc}

In summary, we have investigated the optimization of 2D single-inlet/multi-outlet manifolds, focusing on a square design domain with a fixed volume constraint and covering a wide range of gas rarefaction degrees.
The most notable findings of this study include the following two points:
\begin{enumerate}
\item Wetted-area minimum: As the Knudsen number increases, the wetted area of the optimal manifold first decreases and then increases, with a minimum occurring when  $0.01 \lessapprox\text{Kn}\lessapprox 0.1$, which suggests a ``\emph{small wetted area}'' design principle around this Knudsen number. We attribute this extremum to the onset of velocity slip at the wall induced by rarefaction effects, which weakens the influence of wall friction at higher Knudsen numbers. Notably, this wetted-area minimum does not coincide with the classical Knudsen minimum (occurring near Kn $\approx$ 1), which arises from different physical mechanisms.
\item Inlet--outlet reciprocity: As the Knudsen number increases, the differences in both optimal manifold configuration and MFR caused by exchanging  inlet and outlet conditions diminish progressively. In the free-molecular regime, exchanging the inlet and outlet conditions yields nearly identical optimal configurations and MFRs. This  inlet–outlet reciprocity stems from the homogeneous linear nature of the collisionless Boltzmann equation, which governs free-molecular gas flow. This also implies that the optimal manifold design in this regime is independent of the inlet--outlet pressure ratio. 
\end{enumerate}

In addition, the study reveals several other patterns:
\begin{itemize}
\item At low Knudsen numbers, the optimal channel exhibits smooth curvature at flow direction changes, whereas at high Knudsen numbers, it tends to change direction abruptly and forms straight channels. 
\item Compressibility has a significant influence at low Knudsen numbers, while its effect is relatively minor at high Knudsen numbers.
\item The impact of the flow uniformity constraint on the optimal total MFR primarily depends on the strength of wall friction, thus its effect is most pronounced near the aforementioned wetted-area minimum.
\end{itemize}

We believe this study provides valuable insights for the design of manifolds and piping systems in rarefied environments, and reveals a series of gas flow mechanisms and corresponding optimization principles that emerge at moderate-to-large Knudsen numbers.

\section*{Acknowledgments}
This work is supported by the National Natural Science Foundation of China (12402388). The authors acknowledge the computing resources from the Center for Computational Science and Engineering at the Southern University of Science and Technology.

\bibliographystyle{elsarticle-num} 
\bibliography{2025_manifold}

\begin{thebibliography}{10}
\expandafter\ifx\csname url\endcsname\relax
  \def\url#1{\texttt{#1}}\fi
\expandafter\ifx\csname urlprefix\endcsname\relax\def\urlprefix{URL }\fi
\expandafter\ifx\csname href\endcsname\relax
  \def\href#1#2{#2} \def\path#1{#1}\fi

\bibitem{ceviz2010design}
M.~Ceviz, M.~Ak{\i}n, {Design of a new SI engine intake manifold with variable
  length plenum}, Energy Conversion and Management 51~(11) (2010) 2239--2244.

\bibitem{bassiouny1984flow2}
M.~Bassiouny, H.~Martin, {Flow distribution and pressure drop in plate heat
  exchangers---II Z-type arrangement}, Chemical Engineering Science 39~(4)
  (1984) 701--704.

\bibitem{johnson2012model}
T.~A. Johnson, M.~P. Kanouff, D.~E. Dedrick, G.~H. Evans, S.~W. Jorgensen,
  Model-based design of an automotive-scale, metal hydride hydrogen storage
  system, International Journal of Hydrogen Energy 37~(3) (2012) 2835--2849.

\bibitem{roman2007fully}
G.~T. Roman, R.~T. Kennedy, Fully integrated microfluidic separations systems
  for biochemical analysis, Journal of Chromatography A 1168~(1-2) (2007)
  170--188.

\bibitem{bassiouny1984flow}
M.~K. Bassiouny, H.~Martin, {Flow distribution and pressure drop in plate heat
  exchangers---I U-type arrangement}, Chemical Engineering Science 39~(4)
  (1984) 693--700.

\bibitem{pan2008optimal}
M.~Pan, Y.~Tang, L.~Pan, L.~Lu, Optimal design of complex manifold geometries
  for uniform flow distribution between microchannels, Chemical Engineering
  Journal 137~(2) (2008) 339--346.

\bibitem{emerson2006biomimetic}
D.~R. Emerson, K.~Cie{\'s}licki, X.~Gu, R.~W. Barber, Biomimetic design of
  microfluidic manifolds based on a generalised {M}urray's law, Lab on a Chip
  6~(3) (2006) 447--454.

\bibitem{choi1993effect}
S.~H. Choi, S.~Shin, Y.~I. Cho, The effect of area ratio on the flow
  distribution in liquid cooling module manifolds for electronic packaging,
  International Communications in Heat and Mass Transfer 20~(2) (1993)
  221--234.

\bibitem{delsman2004microchannel}
E.~Delsman, A.~Pierik, M.~De~Croon, G.~Kramer, J.~Schouten, Microchannel plate
  geometry optimization for even flow distribution at high flow rates, Chemical
  Engineering Research and Design 82~(2) (2004) 267--273.

\bibitem{kubo2017level}
S.~Kubo, K.~Yaji, T.~Yamada, K.~Izui, S.~Nishiwaki, A level set-based topology
  optimization method for optimal manifold designs with flow uniformity in
  plate-type microchannel reactors, Structural and Multidisciplinary
  Optimization 55~(4) (2017) 1311--1327.

\bibitem{jensen2018topology}
F.~Jensen, Topology optimization of turbine manifold in the rocket engine
  demonstrator prometheus, Master's thesis, Lule{\aa} University of Technology
  (2018).

\bibitem{bakshi2009euv}
V.~Bakshi, {EUV Lithography}, SPIE press, 2009.

\bibitem{garimella2013simulation}
S.~Garimella, X.~Zhou, Z.~Ouyang, Simulation of rarefied gas flows in
  atmospheric pressure interfaces for mass spectrometry systems, Journal of the
  American Society for Mass Spectrometry 24~(12) (2013) 1890--1899.

\bibitem{ewart2007mass}
T.~Ewart, P.~Perrier, I.~A. Graur, J.~G. M{\'e}olans, Mass flow rate
  measurements in a microchannel, from hydrodynamic to near free molecular
  regimes, Journal of Fluid Mechanics 584 (2007) 337--356.

\bibitem{sato2019topology}
A.~Sato, T.~Yamada, K.~Izui, S.~Nishiwaki, S.~Takata, {A topology optimization
  method in rarefied gas flow problems using the Boltzmann equation}, Journal
  of Computational Physics 395 (2019) 60--84.

\bibitem{caflisch2021adjoint}
R.~Caflisch, D.~Silantyev, Y.~Yang, {Adjoint DSMC for nonlinear Boltzmann
  equation constrained optimization}, Journal of Computational Physics 439
  (2021) 110404.

\bibitem{yuan2024design}
R.~Yuan, L.~Wu, A design optimization method for rarefied and continuum gas
  flows, Journal of Computational Physics 517 (2024) 113366.

\bibitem{bendsoe2003topology}
M.~P. Bends{\o}e, O.~Sigmund, Topology optimization: theory, methods, and
  applications, Springer Science \& Business Media, 2003.

\bibitem{borrvall2003topology}
T.~Borrvall, J.~Petersson, {Topology optimization of fluids in Stokes flow},
  International Journal for Nnumerical Methods in Fluids 41~(1) (2003) 77--107.

\bibitem{bhatnagar1954model}
P.~L. Bhatnagar, E.~P. Gross, M.~Krook, A model for collision processes in
  gases. {I. Small} amplitude processes in charged and neutral one-component
  systems, Physical Review 94~(3) (1954) 511.

\bibitem{Bird1994Molecular}
G.~A. Bird, Molecular gas dynamics and the direct simulation of gas flows,
  Clarendon Press, 1994.

\bibitem{peter2010numerical}
J.~E. Peter, R.~P. Dwight, {Numerical sensitivity analysis for aerodynamic
  optimization: A survey of approaches}, Computers \& Fluids 39~(3) (2010)
  373--391.

\bibitem{yuan2020conservative}
R.~Yuan, C.~Zhong, A conservative implicit scheme for steady state solutions of
  diatomic gas flow in all flow regimes, Computer Physics Communications 247
  (2020) 106972.

\bibitem{yuan2021multi}
R.~Yuan, S.~Liu, C.~Zhong, A multi-prediction implicit scheme for steady state
  solutions of gas flow in all flow regimes, Communications in Nonlinear
  Science and Numerical Simulation 92 (2021) 105470.

\bibitem{svanberg1987method}
K.~Svanberg, The method of moving asymptotes---a new method for structural
  optimization, International Journal for Numerical Methods in Engineering
  24~(2) (1987) 359--373.

\bibitem{svanberg2002class}
K.~Svanberg, A class of globally convergent optimization methods based on
  conservative convex separable approximations, SIAM Journal on Optimization
  12~(2) (2002) 555--573.

\bibitem{johnson2007nLopt}
S.~G. Johnson, The {NLopt} nonlinear-optimization package,
  \url{https://github.com/stevengj/nlopt} (2007).

\bibitem{gersborg2005topology}
A.~Gersborg-Hansen, O.~Sigmund, R.~B. Haber, Topology optimization of channel
  flow problems, Structural and multidisciplinary optimization 30 (2005)
  181--192.

\bibitem{pingen2007topology}
G.~Pingen, A.~Evgrafov, K.~Maute, {Topology optimization of flow domains using
  the lattice Boltzmann method}, Structural and Multidisciplinary Optimization
  34 (2007) 507--524.

\bibitem{akhlaghi2023comprehensive}
H.~Akhlaghi, E.~Roohi, S.~Stefanov, {A comprehensive review on micro-and
  nano-scale gas flow effects: Slip-jump phenomena, Knudsen paradox,
  thermally-driven flows, and Knudsen pumps}, Physics Reports 997 (2023) 1--60.

\bibitem{szalmas2015analysis}
L.~Szalm{\'a}s, T.~Veltzke, J.~Th{\"o}ming, Analysis of the diodic effect of
  flows of rarefied gases in tapered rectangular channels, Vacuum 120 (2015)
  147--154.

\bibitem{graur2016physical}
I.~Graur, J.~M{\'e}olans, P.~Perrier, J.~Th{\"o}ming, T.~Veltzke, A physical
  explanation of the gas flow diode effect, Microfluidics and Nanofluidics
  20~(10) (2016) 145.

\bibitem{hemadri2018liquid}
V.~Hemadri, V.~Duryodhan, A.~Agrawal, Liquid and gas flows in microchannels of
  varying cross section: a comparative analysis of the flow dynamics and design
  perspectives, Microfluidics and Nanofluidics 22~(2) (2018) 19.

\end{thebibliography}

\end{document}